\documentclass[journal]{IEEEtran}
\IEEEoverridecommandlockouts
\usepackage{booktabs}
\usepackage{siunitx}
\usepackage{cite}
\usepackage{amsmath,amssymb,amsfonts}
\usepackage{graphics, framed}
\usepackage{graphicx}
\usepackage{epsfig}
\usepackage{epstopdf}
\usepackage{url}
\usepackage{multimedia}
\usepackage{paralist}
\usepackage[printonlyused]{acronym}
\usepackage{units}
\usepackage{algorithmic}
\usepackage{textcomp}
\usepackage{float}
\usepackage{tabularx}
\usepackage{balance}
\usepackage{subfigure}
\usepackage{xcolor}
\def\BibTeX{{\rm B\kern-.05em{\sc i\kern-.025em b}\kern-.08em
T\kern-.1667em\lower.7ex\hbox{E}\kern-.125emX}}
\usepackage{mathtools}
\usepackage{cuted} 
\usepackage[export]{adjustbox} 

\usepackage{dblfloatfix}
\usepackage{lipsum}
\usepackage{color,soul}
\usepackage{multirow}
\usepackage[overload]{textcase}
\newcommand{\FGR}[1]{Fig.~\ref{#1}}

\newcommand{\SEC}[1]{Section~\ref{#1}}
\newcommand{\TAB}[1]{Table~\ref{#1}}
\newcommand{\EQ}[1]{(\ref{#1})}
\acrodef{5G}[5G]{5\textsuperscript{th}-Generation}
\acrodef{BW}[BW]{bandwidth}
\acrodef{BER}[BER]{bit error rate}
\acrodef{DEM}[DEM]{digital elevation model}
\acrodef{BPSK}[BPSK]{binary phase-shift keying}
\acrodef{CW}[CW]{continuous wave}
\acrodef{WPT}[WPT]{wireless power transmission}
\acrodef{GHz}[GHz]{gigahertz}
\acrodef{GAT}[GAT]{graph attention network}
\acrodef{THz}[THz]{Terahertz}
\acrodef{ISL}[ISL]{inter-satellite link}
\acrodef{RIS}[RIS]{Reconfigurable intelligent surface}
\acrodef{GM}[GM]{Gamma mixture}
\acrodef{PSK}[PSK]{phase shift keying}
\acrodef{QAM}[QAM]{quadrature amplitude modulation}
\acrodef{AWGN}[AWGN]{additive white Gaussian noise}
\acrodef{SNR}[SNR]{signal-to-noise ratio}
\acrodef{AF}[AF]{amplitude-and-forward}
\acrodef{MIMO}[MIMO]{multiple-input multiple-output}
\acrodef{mMIMO}[mMIMO]{massive-multiple-input multiple-output}
\acrodef{SDN}[SDN]{Software-defined network}
\acrodef{SON}[SON]{self-organizing network}
\acrodef{HetNet}[HetNet]{heterogeneous network}
\acrodef{FSO}[FSO]{free-space optics}
\acrodef{UM-MIMO}[UM-MIMO]{ultra-massive-MIMO}
\acrodef{AP}[AP]{access point}
\acrodef{UE}[UE]{user equipment}
\acrodef{NTN}[NTN]{non-terrestrial network}
\acrodef{UAV}[UAV]{unmanned aerial vehicle}
\acrodef{HAPS}[HAPS]{high-altitude platform station}
\acrodef{LEO}[LEO]{low-Earth orbit}
\acrodef{BAN}[BAN]{body area network}
\acrodef{WLAN}[WLAN]{wireless local area network}
\acrodef{QoS}[QoS]{quality of service}
\acrodef{TCS}[TCS]{thermal control system}
\acrodef{QCL}[QCL]{quantum cascade laser}
\acrodef{CMOS}[CMOS]{complementary metal-oxide semiconductor}
\acrodef{V-HetNet}[V-HetNet]{vertical heterogeneous network}
\acrodef{DL}[DL]{deep learning}
\acrodef{DRL}[DRL]{deep reinforcement learning}
\acrodef{EIRP}[EIRP]{effective isotropic radiated power}
\acrodef{FDTD}[FDTD]{Finite-difference time-domain}
\acrodef{FEM}[FEM]{finite element method}
\acrodef{MoM}[MoM]{method of moments}
\acrodef{VNA}[VNA]{vector network analyzer}
\acrodef{CS}[CS]{channel sounder}
\acrodef{CIR}[CIR]{channel impulse response}
\acrodef{CTF}[CTF]{channel transfer function}
\acrodef{DPM}[DPM]{Dirichlet process mixture}
\acrodef{TOA}[TOA]{time of arrival}
\acrodef{GMM}[GMM]{Gaussian mixture model}
\acrodef{OOK}[OOK]{on-off keying}
\acrodef{MLE}[MLE]{maximum likelihood estimation}
\acrodef{LOS}[LOS]{line-of-sight}
\acrodef{NLOS}[NLOS]{non-line-of-sight}
\acrodef{SG}[SG]{signal generator}
\acrodef{SEP}[SEP]{Sun-Earth-probe}
\acrodef{FDSOI}[FDSOI]{fully depleted silicon on insulator}
\acrodef{OpEx}[OpEx]{operational expenditures}
\acrodef{TCO}[TCO]{total cost of ownership}
\acrodef{CapEx}[CapEx]{capital expenditures}
\acrodef{MAC}[MAC]{medium access control}
\acrodef{GEO}[GEO]{geostationary orbit}
\acrodef{SWaP}[SWaP]{size, weight, and power}
\acrodef{EH}[EH]{energy harvesting}
\acrodef{ZE}[ZE]{zero-energy}
\acrodef{IoT}[IoT]{Internet of Things}

\ifCLASSINFOpdf
\else
\fi

\begin{document}
\title{Wireless Power Transmission on Martian Surface for Zero-Energy Devices}

\author{K{\"{u}}r{\c{s}}at~Tekb{\i}y{\i}k,~\IEEEmembership{Graduate Student Member,~IEEE,} Dogay Altinel, Mustafa Cansiz,~\IEEEmembership{Members,~IEEE}  G{\"{u}}ne{\c{s}}~Karabulut~Kurt,~\IEEEmembership{Senior~Member,~IEEE} 

\thanks{K. Tekb{\i}y{\i}k is with the Department of Electronics and Communications Engineering, {\.{I}}stanbul Technical University, {\.{I}}stanbul, Turkey (e-mail: tekbiyik@itu.edu.tr).}
\thanks{Dogay Altinel is with the Department of Electrical and Electronics Engineering, Istanbul Medeniyet University, 34700 Istanbul, Turkey (e-mail: dogay.altinel@medeniyet.edu.tr)}
\thanks{Mustafa Cansiz is with the Department of Electrical and Electronics Engineering, Dicle University, 21280 Diyarbakır, Turkey (e-mail: mustafa.cansiz@dicle.edu.tr)}
\thanks{G. Karabulut Kurt is with the Department of Electrical Engineering and Poly-Grames Research Center, Polytechnique Montr\'eal, Montr\'eal, Canada (e-mail: gunes.kurt@polymtl.ca).}
 }

\IEEEoverridecommandlockouts 

\maketitle

\begin{abstract}
Exploration of the Red Planet is essential on the way through both human colonization and establishing a habitat on the planet. Due to the high costs of space missions, the use of distributed sensor networks has been investigated to make in situ explorations affordable. Along with this, the devices with ultra-low-power receivers, which are called \ac{ZE} devices, can pave the way to further discoveries for the environment of Mars. This study focuses on wireless power transmission to provide the power required by \ac{ZE} devices on the Martian surface. The main motivation of this study is to investigate whether conventional harvesters and communication units can supply the required power for a long distance. The numerical results show that it is possible to deliver power to \ac{ZE} devices without utilizing any sophisticated hardware. In addition, the effects of pointing error and dust storms on harvesting performance are investigated. Comprehensive simulation results reveal that harvester selection and design should be done by considering propagation channel and transmitter characteristics.

\end{abstract}


\IEEEpeerreviewmaketitle
\acresetall

\section{Introduction}

Exploring the Red planet, Mars, which has been going on for half a century, has now evolved to establish a new habitat and has focused on the search for traces of life in the past~\cite{MissionO34:online}. Studies indicate that there were various water sources on Mars in the past~\cite{barnes2020multiple}. Although these results, obtained as a result of extensive measurements and research, increase the excitement for human colonization, the need for much more comprehensive analyzes and exploration continues. Considering the cost of space missions, economical ways of conducting in situ explorations have been investigated for a long time and new methods have been proposed for this purpose. Among these, the use of wireless sensor networks for obtaining data such as temperature, pressure, soil properties related to the Martian environment stands out~\cite{bonafini2020building, alazzam2011thermal}.

Moreover, recent developments in \ac{IoT} devices pave the way for ultra-long battery life due to low power consumption and optimization techniques. Furthermore, incorporating \ac{EH}, it is possible to operate batteryless as proposed in the literature~\cite{lopez2021massive}. Also, it should be stated that the power consumption of the devices reduced to a few nWs. To illustrate, the sensor node proposed in~\cite{lee2012modular} consumes 8.64 nW and 228 pW during data communication with 120 bps and standby mode, respectively. Recently, ultra-low-power receivers with \ac{EH} capability, called \ac{ZE} devices, have been proposed to avoid the need for the replacement of sensor batteries~\cite{haque2020supplemental, alves2021wireless}. In~\cite{taha2021eliminating}, it is shown that \ac{ZE} devices can decode messages with power consumption of less than 120 nW. One more example for \ac{ZE} sensors is given in~\cite{zhang2018near} which proposes a temperature sensor with high accuracy while consuming 80 pW in the worst case. Considering the ultra-low power consumption of the state-of-art sensor nodes, it is possible to extend the operating times of IoT devices to a couple of decades~\cite{portilla2019extreme}. However, it should be stated that providing the necessary power for wireless sensor networks is still an important open issue. \ac{EH} and \ac{WPT} are considered key enablers for ultra-low-power distributed wireless sensor networks~\cite{portilla2019extreme, mahmood2020white, mahmood2020six, sherman2021design}. As a state-of-art study, \cite{dardari2019ultra} proposes a localization system based on batteryless sensor nodes powered by energy harvesting for Mars missions. Due to the low power consumption of \ac{ZE} devices, it is possible to provide the energy needed by the devices by \ac{WPT}. In this context, this study considers \ac{WPT} for \ac{ZE} devices on the Martian surface as illustrated in \FGR{fig:system_model} and provides initial results. In the considered context, \ac{ZE} devices can harvest the required power from a source that can generate relatively high power from energy resources based on nuclear, solar, or etc. The power might be carried by radio-frequency (RF) signals as depicted in \FGR{fig:system_model}.

Before diving into the details of the study, it would be appropriate to summarize the studies on \ac{WPT} in space missions in order to explain the findings of this study. First of all, studies led by Japan Aerospace Exploration Agency (JAXA) were carried out to transmit the energy obtained from solar farms to the Earth~\cite{sasaki2014japan}. Moreover, JAXA developed an experimental system for power transmission to moving rovers~\cite{fuse2011outline, fuse2011microwave}. Besides JAXA, National Aeronautics and Space Administration (NASA) is also working on high-power wireless transmission and is expected to reach TRL 6 by 2028~\cite{936Scott70:online}. In another study~\cite{liu2015multiple}, magnetic resonant coupling enabled by \ac{WPT} for charging distributed magnetic sensors is discussed. In that study, a rover on Mars was selected as the energy source. The RF signal in the high-frequency (HF) band transmitted from the transmitter units on the rover is used for charging the magnetic sensors. Since the aforementioned studies aim to transmit ultra-high power, a long-term development process is needed in order to be used practically. 

On the other hand, as mentioned above, a decrease in the power consumption by the state-of-art sensors can pave the way for low-power WPT. Considering the recent attention on the spatially distributed sensors and data fusion for exploration missions and the open issue on energy sources for distributed nodes, we investigate a possible solution based on WPT to the open issue. In this regard, the contributions of this study can be summarized as:

\begin{enumerate}[{C}1]
    \item To the best knowledge of authors, this study firstly considers \ac{WPT} for \ac{ZE} devices on the Martian surface.
    \item WPT performance is investigated under misalignment fading and dust storms for different harvester designs.
    \item Last but not least, this study shows that the channel characteristics, operating environment, and transmitter capacity must be considered during harvester selection.
\end{enumerate}

The remainder of the paper is organized as follows: \SEC{sec:environment} gives a brief summary on the environmental characteristics of Mars and reasoning for using RF-based harvesting. \SEC{sec:transmission_mars} addresses the mathematical background of the \ac{WPT} system on the Martian surface. In \SEC{sec:results}, the harvested power depending on transmission power, distance, dust storms, and pointing error is discussed over numerical results. \SEC{sec:open_issues} addresses the open issues of this study and provides direction for future work. Finally, \SEC{sec:conclusion} concludes the paper.

\begin{figure}[!t]
    \centering
    \includegraphics[width=0.8\linewidth, page = 3]{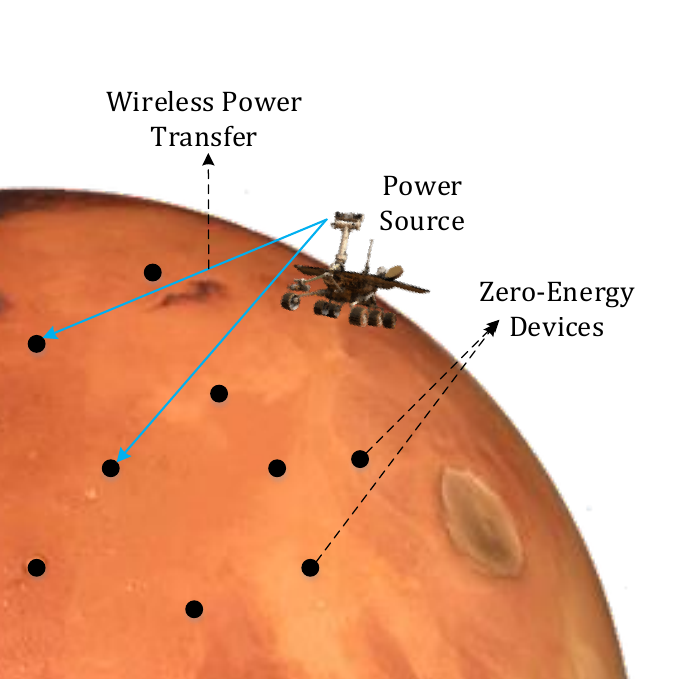}
    \caption{Illustration for wireless transmission from an energy source to remote \ac{ZE} devices.}
    \label{fig:system_model}
\end{figure}

\section{Environmental Conditions and Energy on Mars}\label{sec:environment}

Martian atmosphere consists mainly of carbon dioxide and the atmospheric pressure of the Mars slightly lower than 1\% of the atmospheric pressure at the surface of Earth~\cite{landis2000solar}. This atmosphere consists of approximately 95.5\% carbon dioxide, 2.7\% nitrogen, 1.6\% argon, 0.15\% oxygen and other gases\cite{kaplan1988environment}. There are suspended dust particles in the Martian atmosphere, and based on local and global storms, the amounts of these dust particles change daily and seasonally~\cite{appelbaum1990solar}. In each Martian year, global dust storms may occur one or two times on occasion in planetary scale. The duration of these global dust storms may change from 35 to 70 days or more. Compared to the global dust storms, the intensity of local dust storms is lower and they disappear in a few days or less~\cite{appelbaum1990solar}. A Martian year is 1.88 terrestrial years and a Martian day is 24.62 terrestrial hours.

Dust particles in the Martian atmosphere reduce the solar intensity at the surface of the Mars. The amount of dust particles in the atmosphere is measured by optical depth which have no unit. Based on the latitude, season, and dust storms, the value of optical depth can change from less than 0.4 to more than 4~\cite{landis2000solar}. The dust particles in the atmosphere affect the solar spectrum and intensity at the surface of the Mars. Dust particles scatter in the red end of the solar spectrum and absorb in the blue end~\cite{landis2000solar}. The effects of dust particles on the solar intensity on the surface have been investigated by the various researchers~\cite{haberle1993atmospheric}. On the Pathfinder mission, performance of the solar cells was also analyzed. Pathfinder was designed to deliver an instrumented lander and the first ever robotic rover to the Martian surface and accomplished that purpose. Pathfinder was landed on the surface of Mars on July 4, 1997.

In terms of the air temperature, Mars is a very cold planet compared to the Earth. During a Martian year, the temperatures of the air at a height of 1.6 meters above the surface were acquired by Viking Lander 1 and 2 (including global and local dust storms). NASA's Viking Project became the first United States mission to land a spacecraft safely on the Martian surface and send photographs of the surface. Viking Lander 1 and 2 were landed on the surface of Mars July 20, 1976 and September 3, 1976, respectively. The surface temperature of the Mars varies from 130 °K to 300 °K (with an average of 215 °K)~\cite{kaplan1988environment}. Low temperature may affect the performance of electronic devices. Due to the thin atmosphere of the Mars, wind speeds are averagely not very high and wind force is not strength. At the Viking Lander 2, the average wind speed was measured as approximately 2 m/s~\cite{hess1977meteorological}. Besides, the wind speed was measured over 17 m/s less than 1\% of the observation time.

Solar and nuclear energy systems can be used to operate the spacecraft on Mars missions. Each energy system has its own advantages and disadvantages. Mars has quite different environmental conditions from the Earth, and these environmental conditions affect the performance of solar cell array. The main factors impacting the performance of solar cell array at the surface of the Mars can be listed as follows~\cite{landis2004mars}:
\begin{itemize}
    \item Low solar intensity (due to further distance of Mars from Sun compared to Earth)
    \item Suspended dust particles in the Martian atmosphere (these particles modify the solar spectrum and intensity)
    \item Low operating temperature
    \item Deposition of dust particles on the solar cell array
\end{itemize}
On the other hand, the power generated by the solar cell arrays must be stored in an energy storage system for use at the Martian nights. Sodium-sulfur, secondary lithium batteries, silver-zinc and hydrogen-oxygen alkaline regenerative fuel cells were considered as advanced energy storage systems. Because of the high specific energy density, the hydrogen-oxygen alkaline regenerative fuel cells were selected as advanced energy storage system for the long storage periods~\cite{mckissock1990solar}.

Nuclear energy can be used when the power produced by the solar cell arrays is not sufficient. Nuclear energy systems have many advantages such as ease of packaging and compactness. Besides, nuclear energy systems are insensitive to the environmental conditions, and can generate power in the absence of sunlight at the Martian nights~\cite{kerslake1999solar}. Despite its many advantages, this energy system can pollute the environment. 

The mission of Mars rover Perseverance is to detect the signs of life and collect the soil and rock samples for sending to Earth. Perseverance was successfully landed on the surface of Mars Feb. 18, 2021. The electric power for Perseverance is provided by a system called a multi-mission radioisotope thermoelectric generator. Multi-mission radioisotope thermoelectric generator is essentially a nuclear battery and it uses the heat from the radioactive decay of plutonium to generate electric power.

As given in~\cite{alves2021wireless}, the main outstanding feature of RF EH is to have low hardware complexity. On the other hand, solar EH can provide more energy but it should be noted that this comparison is given for the Earth and the longer distance between Mars and Sun results in lower solar flux density. To explain briefly, the amount of solar radiation reaching the Earth is inversely proportional to the square of its distance from the Sun. The Sun-Mars mean distance is $1.5236915$ AU; therefore, the amount of incident solar power on Mars is almost 43\% of the amount of power on the Earth~\cite{appelbaum1990solar}. It is worth saying that this comparison does not even include any dust storms. On the other hand, harvesting from other energy sources requires sophisticated hardware even though their energy provisions are quite high. Thanks to reducing the power requirement of sensor nodes, RF EH can supply sufficient energy to sensor nodes without using any sophisticated hardware. Although it is possible to use various energy sources on Mars, novel methods are needed to meet the energy needs of the sensor networks that are moving and distributed for discovery missions. For this aim, \ac{WPT} is considered as a promising solution. In general, preliminary studies have been carried out for the use of laser beams~\cite{ortabasi2006powersphere} and RF waves~\cite{iwashimizu2014study} for \ac{WPT}. The weight, size, mass, and limited operation temperature of the laser systems increase the space mission costs. Moreover, laser systems are still considered immature~\cite{xie2013wireless}. Therefore, RF-based \ac{WPT} is discussed in this study.

\section{Wireless Transmission on Martian Surface}\label{sec:transmission_mars}

This section provides basic information about the propagation medium on the Martian surface and the mathematical background on the impact of environmental factors on \ac{WPT}.

\subsection{RF Path Loss Modeling}\label{sec:path_loss}

Although studies on propagation modeling for the Martian environment are limited and immature, some recent findings on this issue provide information about RF propagation on Mars. In early studies on propagation models for Martian surface such as~\cite{chukkala2005simulation, daga2007terrain, del2009ieee802}, the proposed propagation models are mainly based on terrestrial assumptions. And therefore there is a need for comprehensive analysis of RF propagation on the Martian surface and atmosphere based on appropriate approaches and assumptions. A recent study~\cite{bonafini2020evaluation} presented realistic RF propagation models with 3D ray tracing based on high resolution \ac{DEM} of the Martian surface. The employed \ac{DEM} shows Gale Crater which is considered a dry lake. This region constitutes an important pillar of the search for life on Mars. The main reason behind this is that Gale Crater has shown strong indications that there was water on Mars in the past. As it will be remembered, NASA's Curiosity spacecraft also landed in this region in August 2012 and collected data about the geology and environment of the region~\cite{wray2013gale, hassler2014mars, mahaffy2013abundance}.

The received signal at the input of harvester can be defined as follows:
\begin{align}
    y(t) = \frac{1}{\sqrt{PL_{tot}}}h(t)m(t)x(t),
\end{align}
where $x(t)$, $h(t)$, and $m(t)$ denote the transmitted signal with transmit power of $P_{TX}$, small-scale fading, and misalignment fading, respectively. $PL_{tot}$ stands for the total path loss including free-space path loss with shadowing and attenuation due to dust storms. 

First, we will discuss on large-scale fading. In the proposed propagation model~\cite{bonafini2020evaluation}, the generic path loss and log-normal shadowing are utilized with the new parameters which have been obtained through 3D ray tracing over \ac{DEM}. Therefore, we employ log-distance path loss model in this study. The log-distance path loss model is given as follows:
\begin{equation}
    PL = 10\alpha\mathrm{log}(K) + \chi \;\;\; \mathrm{(dB)},
\end{equation}
where $\alpha$ is the path loss exponent. $\chi\sim \mathcal{N}\left(0, \sigma\right)$ denotes zero mean shadow fading in dB. $K$ is free space path loss in Watt given as follows:
\begin{equation}
K =\frac{4 \pi d}{\lambda},
\end{equation}
where $\lambda$ and $d$ denote wavelength of the emitted signal and the distance between the source and harvesting device, respectively. In~\cite{bonafini2020evaluation}, the path loss exponent and shadowing values are found (2.12, 11.41) and (2.37, 13.26) for two different areas (i.e., Area 1 and Area 2) in Gale Crater, respectively. Considering the values for Area 2, it is seen that the propagation environment is lossier and the number of multipath is higher compared to Area 1. As stated in~\cite{bonafini2020evaluation}, Area 2 is rocky whereas Area 1 has a flat environment.

Second, dust storms can heavily affect radio propagation on Martian terrain. Therefore, we adopted the attenuation owing to dust storms in the total path loss model. The attenuation due to dust storms is modeled as follows~\cite{goldhirsh1982parameter}:
\begin{equation}
P_{DS}=\frac{1.029 \cdot 10^{3} \mathrm{Im(\varepsilon)}}{\lambda\left[\left(\mathrm{Re(\varepsilon)}+2\right)^{2}+\mathrm{Im(\varepsilon)}^2\right]} N_{T} \rho_p^{-3}d \;\; \mathrm{(dB)},
\end{equation}
where $\varepsilon$ is the dielectric permittivity of dust particles and it is $4.56 + i0.251$ at 2.45 GHz~\cite{sacchi2019lte}. Also, $\mathrm{Re(\cdot)}$ and $\mathrm{Im(\cdot)}$ denote the real and imaginary part of a complex number, respectively. $N_T$ is particle density which means the total number of particles in unit volume. $\rho_p$ is also mean particle radius. It should be noted that $d$ is the propagation distance in meter.

The received power at the input of harvester at $t$ can be given as follows:
\begin{align}
    P_{RX} = P_{TX} - H - M - PL - P_{DS} + G_T + G_R  \;\; \mathrm{(dB)}, 
\end{align}
where $H = 20\log(h(t))$ and $M = 20\log(m(t))$. $G_T$ and $G_R$ stand for the transmitter and receiver antenna gains, respectively.

\subsection{Misalignment Fading}\label{sec:misalignment}

\begin{figure}[!t]
    \centering
    \includegraphics[width=0.8\linewidth, page = 4]{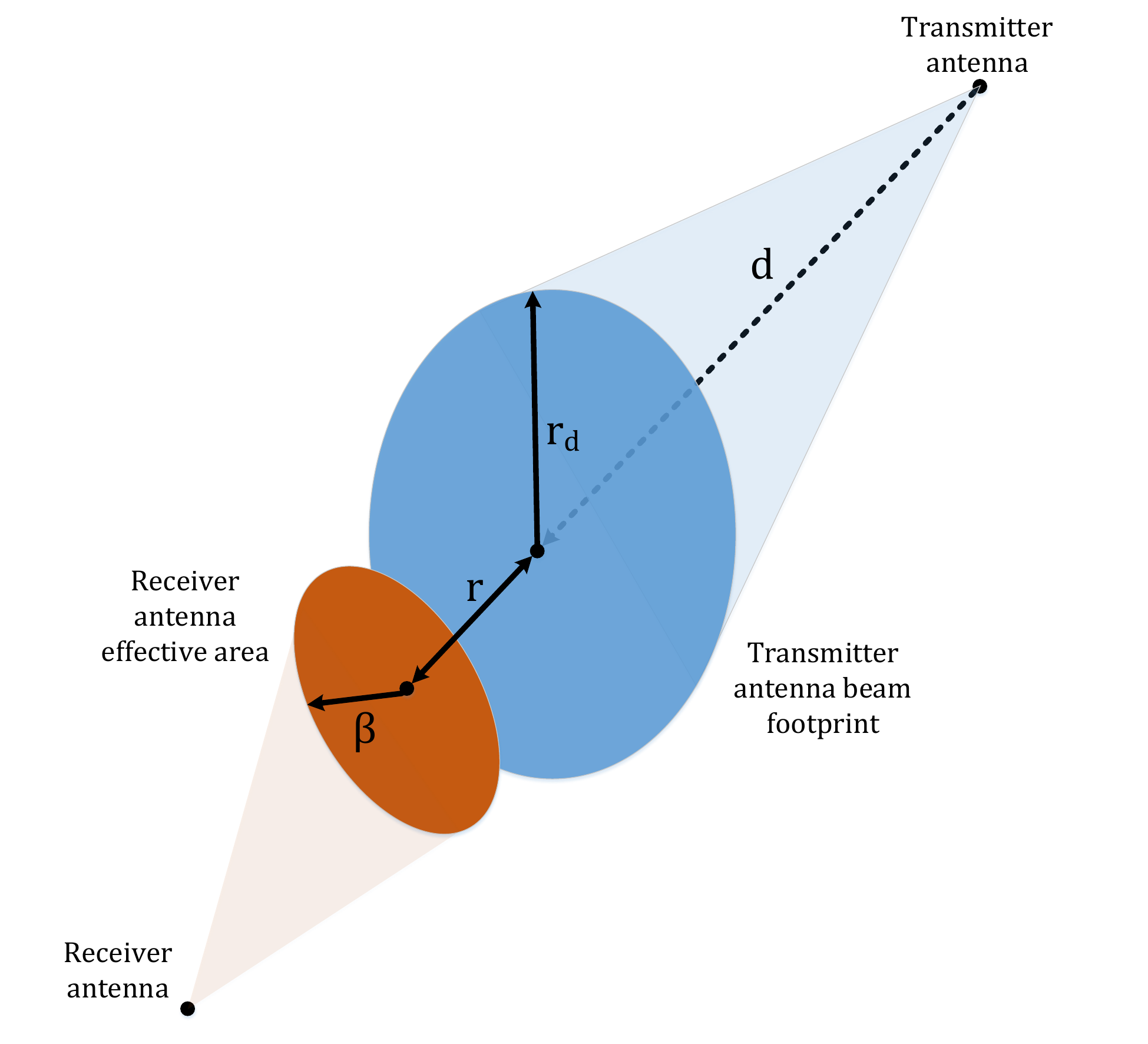}
    \caption{Illustration of the pointing error, $r$, between receiver antenna with effective beam aperture radius, $\beta$, and the transmitter antenna with beam waist, $r_d$ at distance $d$.}
    \label{fig:misalignment}
\end{figure}

In the previous section, we mentioned the misalignment fading without diving into details. However, it is required to give some preliminary details to understand the numerical results. Thus, this section is devoted to giving some preliminaries on the misalignment fading. 

A proper alignment between the source and receiver antennas is essential to receive the power required for operation of \ac{ZE} devices. Because of low-complex hardware and computation capacity of \ac{ZE} devices, proper beam alignment might not be satisfied. Therefore, some alignment errors might be expected. The misalignment fading at a time $t$ can be modeled as a random variable. The remainder of this section follows the steps to obtain the distribution for the random variable.  

First, we assume that the beams are circular. As depicted in~\FGR{fig:misalignment}, for two beams with radial distance r between their centers, the misalignment coefficient, $m = m(t)$, is given as follows~\cite{farid_outage_2007}: 
\begin{align}
    m(r ; d) \approx A_{0} \exp \left(-\frac{2 r^{2}}{w_{e q}^{2}}\right),
    \label{eq:zeta}
\end{align} 
where $w_{eq}$ is the equivalent beamwidth. $A_{0}$ shows the power fraction for aligned antenna pair (i.e., $r = 0$) and it is defined as follows:
\begin{align}
A_{0}= & \left[\operatorname{erf}\left(\frac{\sqrt{\pi}\beta}{\sqrt{2}r_d}\right)\right]^{2},
\end{align}
where $\operatorname{erf}(\cdot)$ and $r_d$ are the error function and the beam waist at distance $d$, respectively. Also, the radius of the receiver antenna's effective area is denoted by $\beta$. The displacement error in two axis can be modeled by identical Gaussian distribution; thus, the radial distance, $r$, follows Rayleigh distribution given by
\begin{align}
    f_{r}(r)=\frac{r}{\sigma_{s}^{2}} \exp \left(-\frac{r^{2}}{2 \sigma_{s}^{2}}\right), \quad r>0,
    \label{eq:rayleigh}
\end{align}
where $\sigma_{s}^{2}$ denotes jitter variance. Utilizing \EQ{eq:zeta} and \EQ{eq:rayleigh} jointly exhibits the misalignment fading with the following distribution:  
\begin{align}
    f_{m}(\zeta) = \frac{\gamma^2}{A_{0}^{\gamma^2}}\zeta^{\gamma^2-1}, \, 0\leq \zeta\leq A_{0},
    \label{eq:misalignment_distribution}
\end{align}
where $\gamma = \frac{w_{eq}^2}{2 \sigma_{s}^2}$~\cite{boulogeorgos_error_2020}. By the analysis given above, it is shown that the jitter variance and the aperture size appear as a crucial factor on misalignment fading and for the harvested power as well.

\subsection{Harvester Efficiency}

\begin{figure}[!t]
    \centering
    \includegraphics[width = \linewidth]{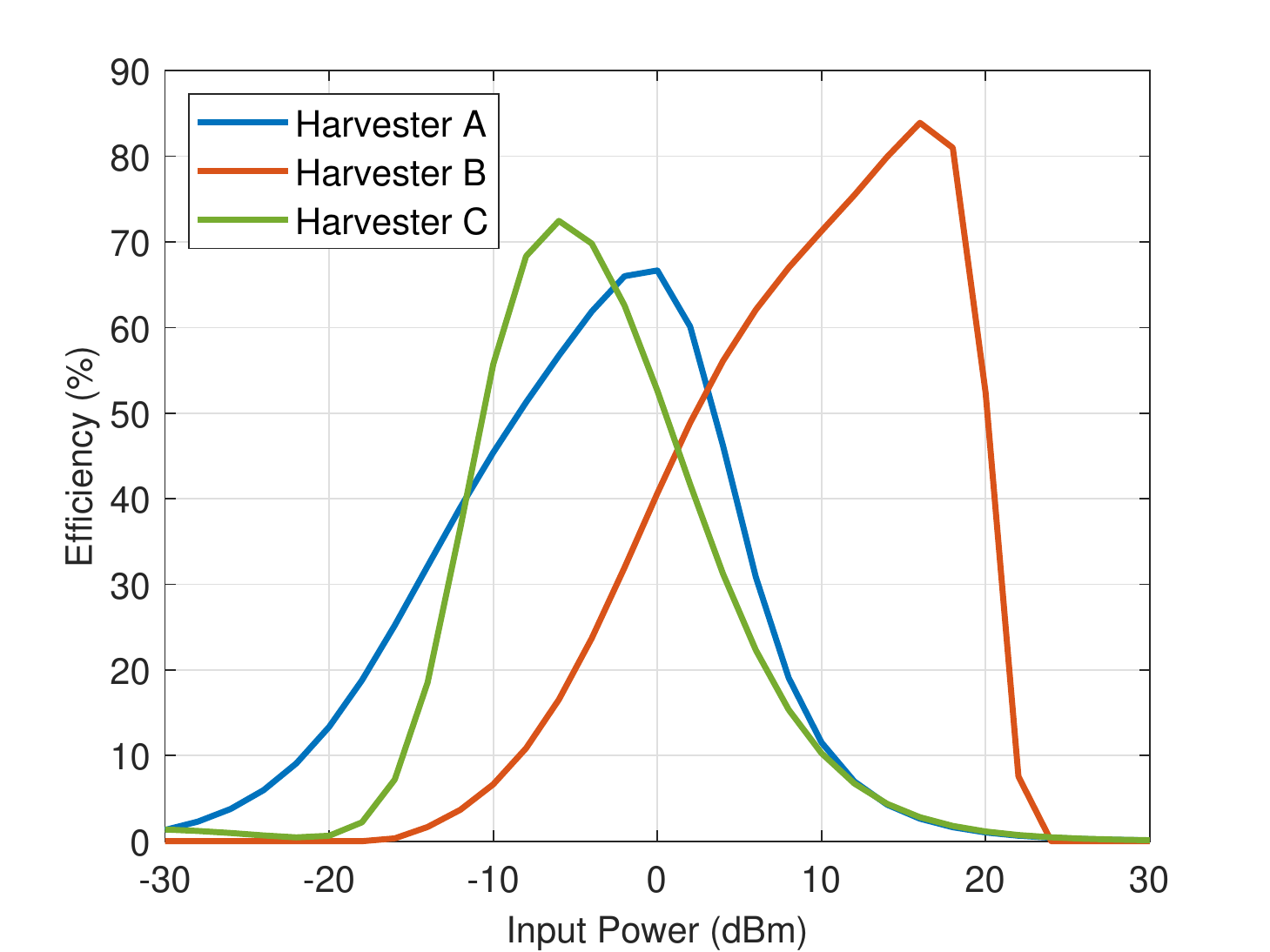}
    \caption{Harvester power conversion efficiency versus incident power for different harvester types.}
    \label{fig:harvester_eff}
\end{figure}

Although harvesters are at the endpoint of energy transmission systems, they are essential for receiving and storing energy~\cite{cansiz2019efficiency}. It should be noted here that due to the subject of this study, only RF energy harvesters are discussed. The seminal works on energy-harvesting wireless systems~\cite{ju2013throughput} consider the harvesters as linear devices whose efficiency is independent of the input power. However, the practical experiments denote that harvesters are nonlinear devices and conversion efficiency is a nonlinear function of the input power~\cite{chen2016new}. As the efficiency of the harvesters has a direct effect on the amount of harvested power, and many studies have been carried out in recent years to increase the efficiency of the harvesters~\cite{valenta2014harvesting, clerckx2018fundamentals}.

As stated above, the harvester efficiency is practically modeled as a nonlinear function of input power. In this regard, several models have been proposed, but the heuristic model is reported with the smallest fitting error~\cite{chen2016new}. Another reason behind using this model is that it allows modeling in a wide scope of input power. The energy conversion efficiency of a harvester is defined with the heuristic model as follows:
\begin{equation}
\eta[P_{RX}]=\frac{a_{2} P_{RX}^{2}+a_{1} P_{RX}+a_{0}}{P_{RX}^{3}+b_{2} P_{RX}^{2}+b_{1} P_{RX}+b_{0}},
\label{eq:heuristic_model}
\end{equation}
where $P_{RX}$ denotes the input power in mW. By employing this model, it is possible to analyze harvesters designed for different input power levels. For example, for infinitely small input power, the efficiency is limited by the $a_{0}/b_{0}$ term, for very large input values, the efficiency is on the order of $1/P_{RX}$. As a result, the harvested power, $P_{h}$, would be given as follows:
\begin{align}
    P_{h} &= P_{RX} \times \eta[P_{RX}], \nonumber \\ 
    & = \frac{a_{2} P_{RX}^{3}+a_{1} P_{RX}^2+a_{0}P_{RX}}{P_{RX}^{3}+b_{2} P_{RX}^{2}+b_{1} P_{RX}+b_{0}}.
\end{align}

In this study, we utilize three different harvesters with different input power levels to investigate the amount of harvested power under several circumstances. Our motivation in selecting harvesters is to cover a wide scope of input power since the input power would be affected by the channel conditions and misalignment. As detailed in \SEC{sec:path_loss}, the received power fluctuates in a wide region due to a relatively high shadowing effect in rocky areas. Hence, the harvester to be used on Mars requires to support a wide input power region. Furthermore, efficiency is another key factor in harvester selection in this study. But, it should be noted that the harvesters pose a trade-off between wide input range and energy conversion efficiency. Since we are aiming to provide an end-to-end analysis on power transfer on Martian surface for \ac{ZE} devices, we need to choose each element of analysis in harmony. Therefore, the utilized harvesters can operate at $2.45$ GHz since channel modeling studies have been focused on that frequency band. To the best knowledge of the authors, the following harvesters seem to comply with the specified conditions: Harvester A~\cite{franciscatto2013high}, Harvester B~\cite{zhang2016high}, and Harvester C~\cite{lau2020deep}. The first two are based on discrete components (e.g., Schottky diode) and the latter is fabricated in CMOS technology. It is worth mentioning the main difference between the two approaches in short. The main advantage of using discrete components in harvester design is to have low loss feature compared to CMOS-based harvesters. Low resistivity silicon substrate in CMOS process induces low Q-factor while the discrete components' Q-factor is quite high, which conduces efficient energy storage~\cite{ramalingam2021advancement}. Because of the promoting features of the Schottky diode such as low forward voltage drop, low power consumption, high switching, and low loss, it stands out in the discrete components~\cite{liu2021research}. However, the size of discrete harvesters is larger compared to CMOS-based architectures.

As depicted in~\FGR{fig:harvester_eff}, the harvesters operate at different incident power range with different efficiency levels, which provides a holistic analysis. By utilizing their measurement data, we employed curve fitting to model the harvesters input power-efficiency relation by the heuristic model given in~\EQ{eq:heuristic_model}. The model parameters are given in~\TAB{tab:model_parameters} for each harvester. 

\begin{table}[!t]
\caption{Model parameters for energy conversion efficiencies of three harvesters investigated throughout this study.}
\resizebox{\linewidth}{!}{
\begin{tabular}{ccccccc}
\toprule \toprule
\multicolumn{1}{l}{\textbf{Harvester}} & $a_{2}$ & $a_{1}$ & $a_{0}$ & $b_{2}$ & $b_{1}$ & $b_{0}$  \\ \toprule
\textbf{A}~\cite{franciscatto2013high}                             &    100.1 & 181.2 & -4.43e-2        &  -6.74e-2 & 3.185 & 10.1e-2  \\ 
\textbf{B}~\cite{zhang2016high}                              &   -5.28e3  & 9.46e5 & -2.04e4   &  -150.6 & 1.292e4 & 9874          \\ 
\textbf{C}~\cite{lau2020deep}                             &   114.6 & -1.613 & 7.66e-3    & 1.133 & 9.84e-3 & 4.5e-3       \\ \bottomrule \bottomrule 
\end{tabular}
}
\label{tab:model_parameters}
\end{table}

\section{Numerical Results and Discussions}\label{sec:results}
In this section, \ac{WPT} on the Martian surface is analyzed in different regions of Gale Crater and under different environmental effects. In all analyzes, the operating frequency was set to 2.45 GHz. This is because - as noted above - the channel modeling studies have generally focused on this band. In addition, while the transmitter antenna gain, $G_T$, is 28 dB, receiver antenna gain, $G_R$, is chosen as 0 dB, assuming that there is no antenna gain due to the simple structure of the receivers. The dust permittivity, $\varepsilon$, is $4.56 + i0.251$ at 2.45 GHz~\cite{sacchi2019lte}. Unless otherwise is stated, the beam waist, $r_d$, the distance between harvesters and the energy source, and the transmitted power are $7\lambda$, 50 m, and 10 W, respectively. Moreover, the path loss exponent and shadowing effect were selected according to propagation medium which is detailed in~\SEC{sec:path_loss}.

\begin{figure}[!t]
    \centering
    \includegraphics[width=\linewidth]{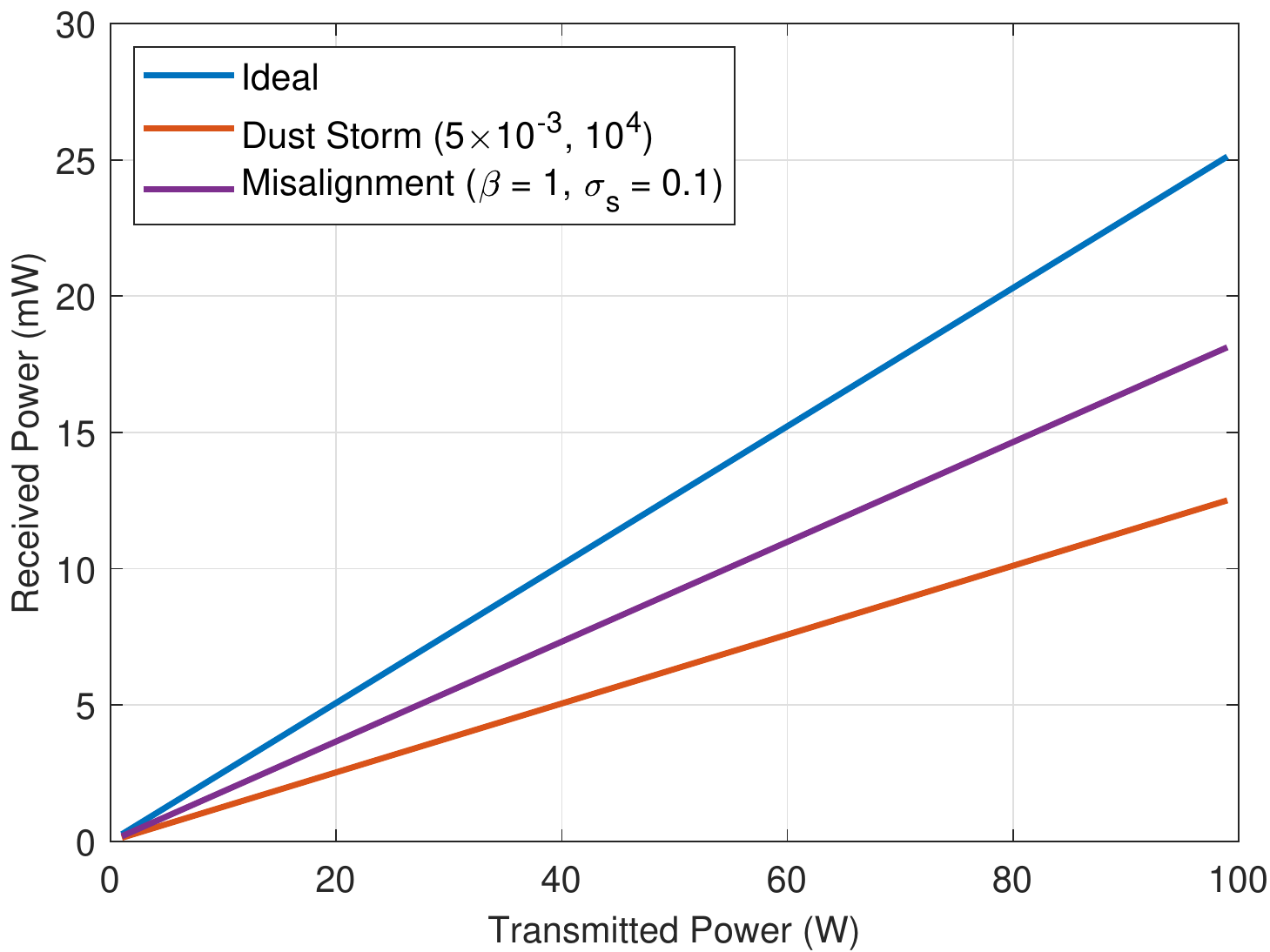}
    \caption{The incident power with respect to transmit power when the propagation distance is 50 m.}
    \label{fig:rx_pow_vs_tx_pow}
\end{figure}

\begin{figure*}[!t]
\centering
\subfigure[]{
\label{fig:tx_pow_50m_dust_0_misalignment_0_area_1}
\includegraphics[width= 0.45\linewidth]{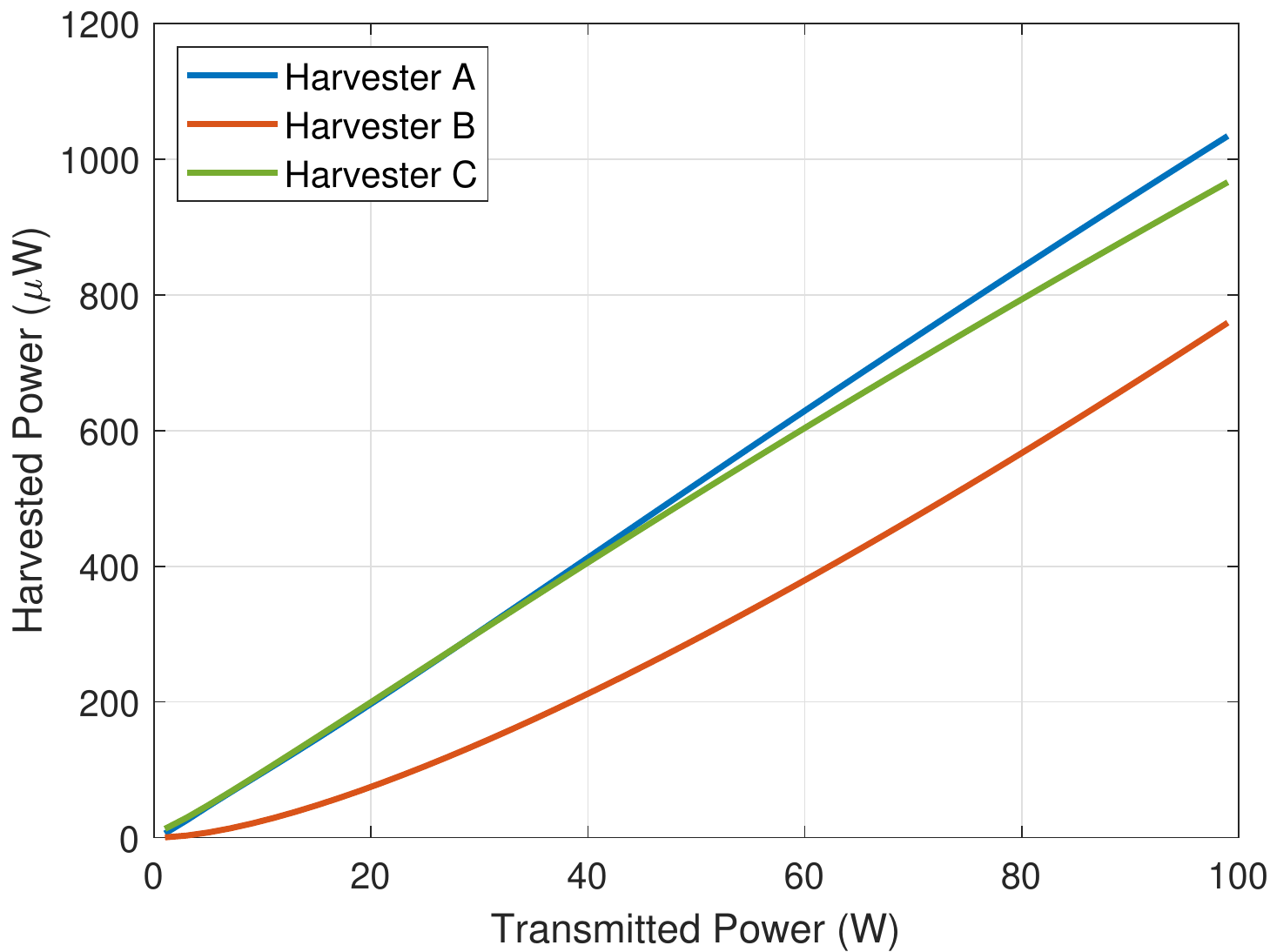}}
\subfigure[]{
\label{fig:tx_pow_50m_dust_0_misalignment_0_area_2}
\includegraphics[width= 0.45\linewidth]{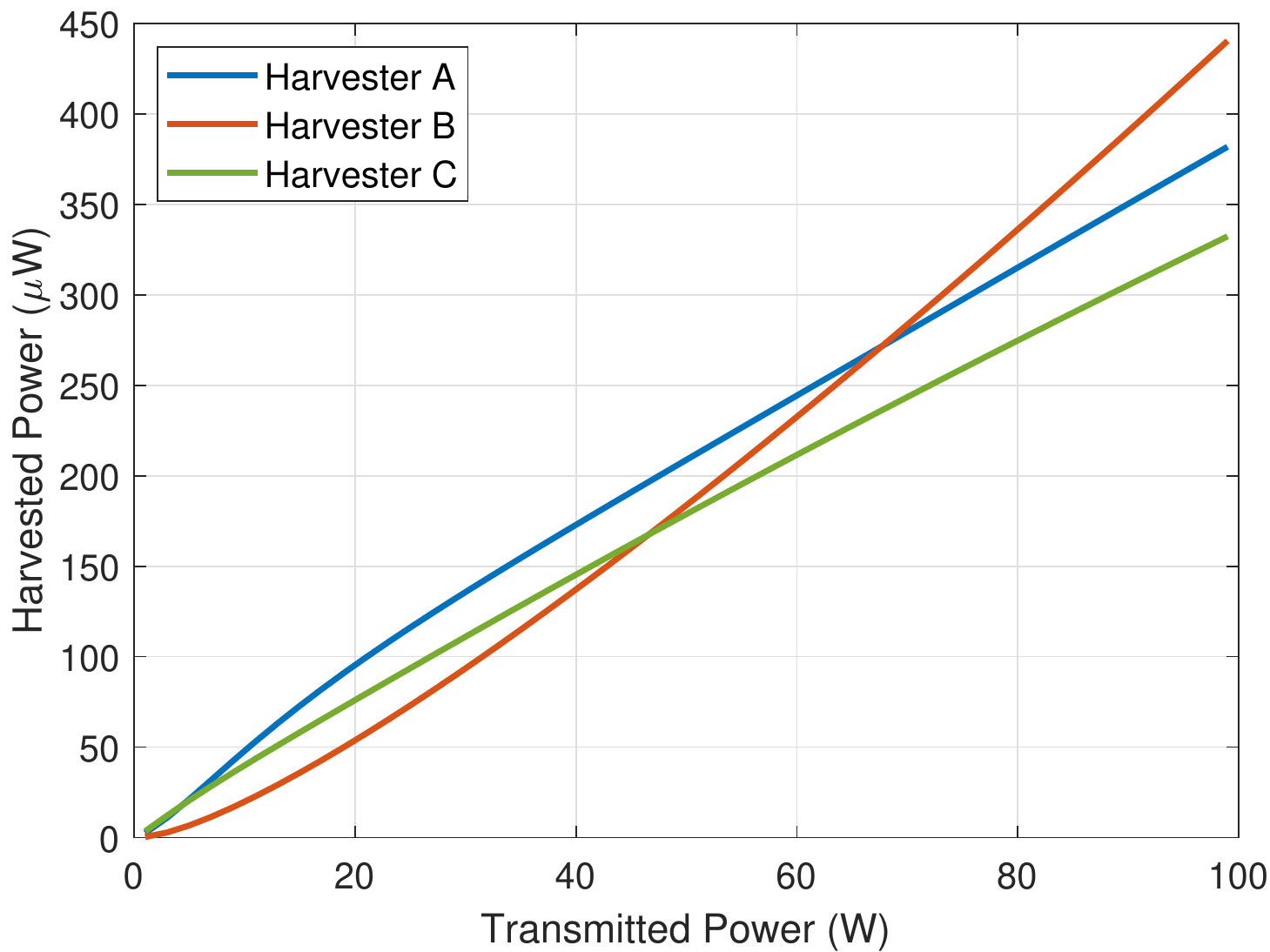}}
\caption{The harvested power by three different harvesters in (a) Gale Crater Area 1 and (b) Gale Crater Area 2 versus varying transmit power when propagation distance is 50 m. Due to high path loss and shadowing in Area 2, the harvested power is low.}
\label{fig:tx_pow_50m_dust_0_misalignment_0}
\end{figure*}

Since we focus directly on how the harvested power changes under various conditions in the rest of this section, it is convenient to give the relationship between the received power and the transmitted power in \FGR{fig:rx_pow_vs_tx_pow} as an insight for the following results. It is worth noting that \FGR{fig:rx_pow_vs_tx_pow} shows the behavior under different conditions, which we describe in detail below, apart from the ideal situation. As expected, there is a linear relationship between received power and transmitted power. However, it should not be forgotten that the harvested power exhibits nonlinear behavior. It can be seen in \FGR{fig:rx_pow_vs_tx_pow} that a very small pointing error causes serious energy loss. On the other hand, even in intense dust storms, a maximum of 50\% loss is experienced in the amount of power taken compared to the ideal situation. This shows the benefit of RF-based transmission compared to optical transmission. A detailed discussion of ideal and practical cases particularly shown is given later in this section.

It has been stated above that harvesters are nonlinear devices. For this reason, the relationship between the harvested power and the transmitted power is also nonlinear. First, to examine this nonlinear relationship, under ideal conditions with perfectly aligned antennas and the absence of dust storms, we investigate how the harvested power by different harvesters versus varying transmit power changes. The power of the transmitted signal ranges from 1 to 100 W. \FGR{fig:tx_pow_50m_dust_0_misalignment_0_area_1} shows that Harvester A and C can provide 200 $\mu$W when the transmitted power is 20 W. However, the incident power lies in the region of Harvester B where the efficiency is low. Furthermore, it is observed that A and C operate in their linear region since the incident power changes in between -10 and 0 dBm. In Area~2, which is more challenging in terms of signal transmission, the incident power is much lower and is usually outside the working area of the harvesters. Although it is seen in~\FGR{fig:tx_pow_50m_dust_0_misalignment_0_area_2} that the harvested power is sufficient for \ac{ZE} devices, it is more appropriate to consider novel harvesters for this region.

\begin{figure*}[!t]
\centering
\subfigure[]{
\label{fig:distance_10W__dust_0_misalignment_0_area_1}
\includegraphics[width= 0.45\linewidth]{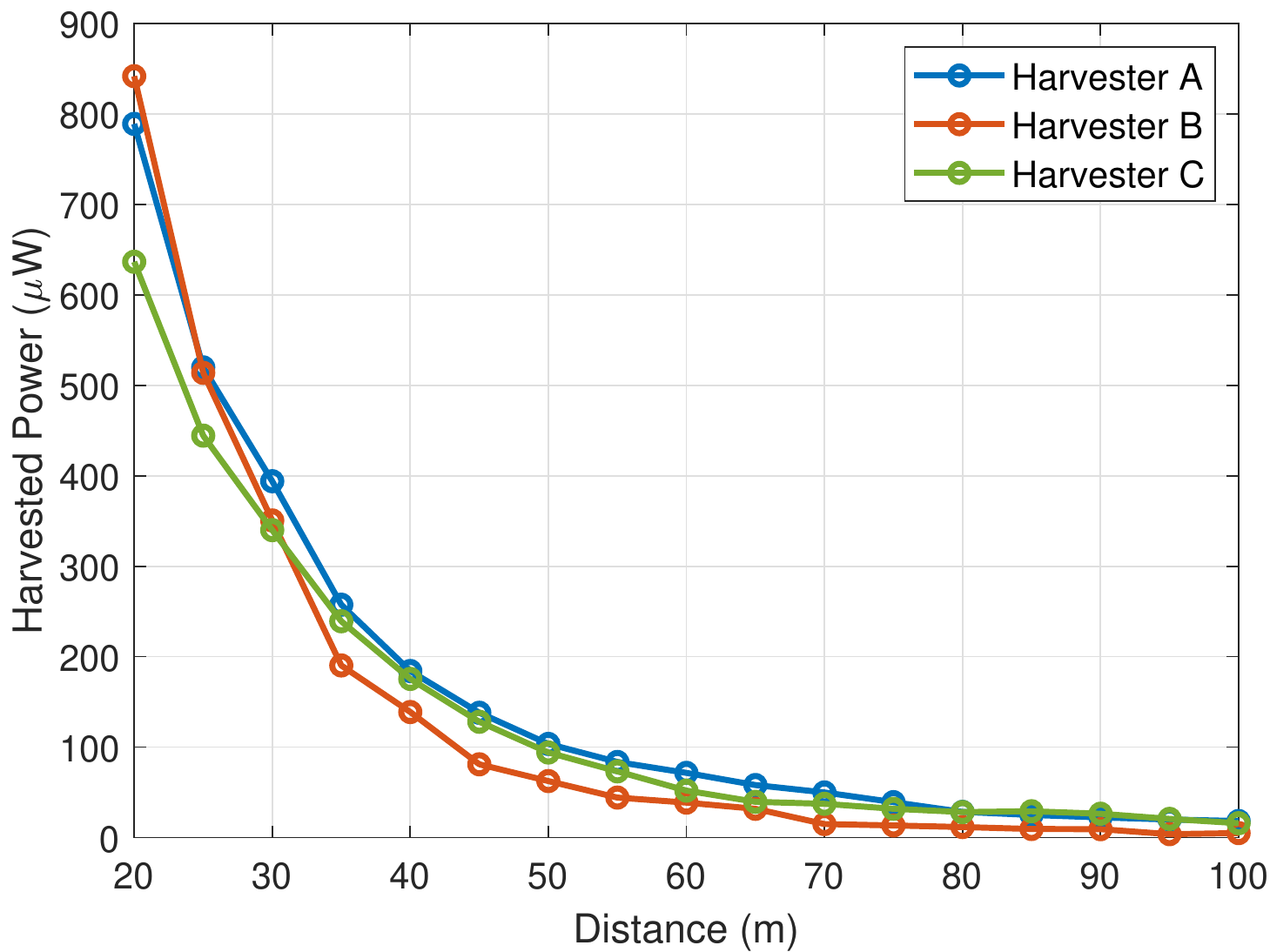}}
\subfigure[]{
\label{fig:distance_10W__dust_0_misalignment_0_area_2}
\includegraphics[width= 0.45\linewidth]{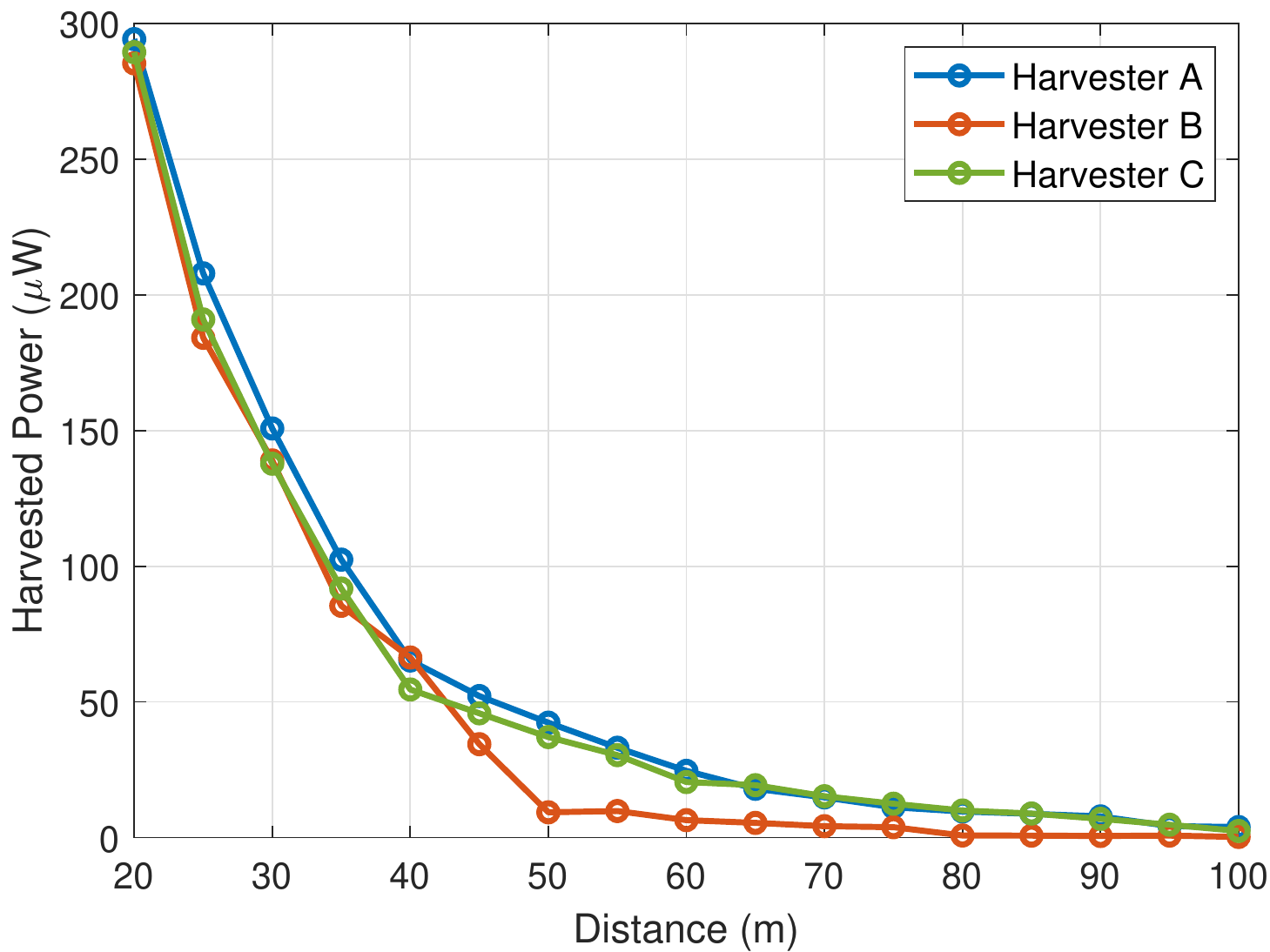}}
\caption{The harvested power by three different harvesters in (a) Gale Crater Area 1 and (b) Gale Crater Area 2 versus propagation distance when the transmit power is 10 W.}
\label{fig:distance_10W__dust_0_misalignment_0}
\end{figure*}

By investigating the amount of energy harvested depending on the transmission distance as well as the transmission power, an insight can be created about how often the resources should be placed. For this aim, the variation of the harvested power depending on the distance was investigated under constant transmit power by considering the two regions of Gale Crater separately. The transmit power is selected as 10 W and the environmental conditions are assumed ideal. As seen in \FGR{fig:distance_10W__dust_0_misalignment_0_area_1}, the harvested power remains above 50 $\mu$W up to 70 m; however, the power decreases and gets meaningless from the point of practical usage for longer distances. It is observed that using Harvester A up to 30 m would be more efficient. If the distance between source and destination nodes is planned to be longer, the harvester selection requires more attention. It is worth saying that the planning strategies for the source deployments should be addressed in further studies. On the other hand, numerical results regarding Area 2 are plotted in \FGR{fig:distance_10W__dust_0_misalignment_0_area_2}. The harvested power is above 50 $\mu$W up to 40 m distance in Area 2. It reveals that energy sources should be placed more frequently within this region. Up to this point, we considered WPT under ideal conditions. Therefore, the difference in the performance of the three harvesters may not be well observed. It should also be noted that ZE devices operate in low power region; thus, a small difference between harvested power by the harvesters can substantially change the efficiency of operation. The rest of this section is devoted to understanding the behavior of harvested power under some practical conditions.

\begin{figure*}[!t]
\centering
\subfigure[]{
\label{fig:particle_density_10W_50m_1e-4_5e-3_dust_1_misalignment_0_area_1}
\includegraphics[width= 0.45\linewidth]{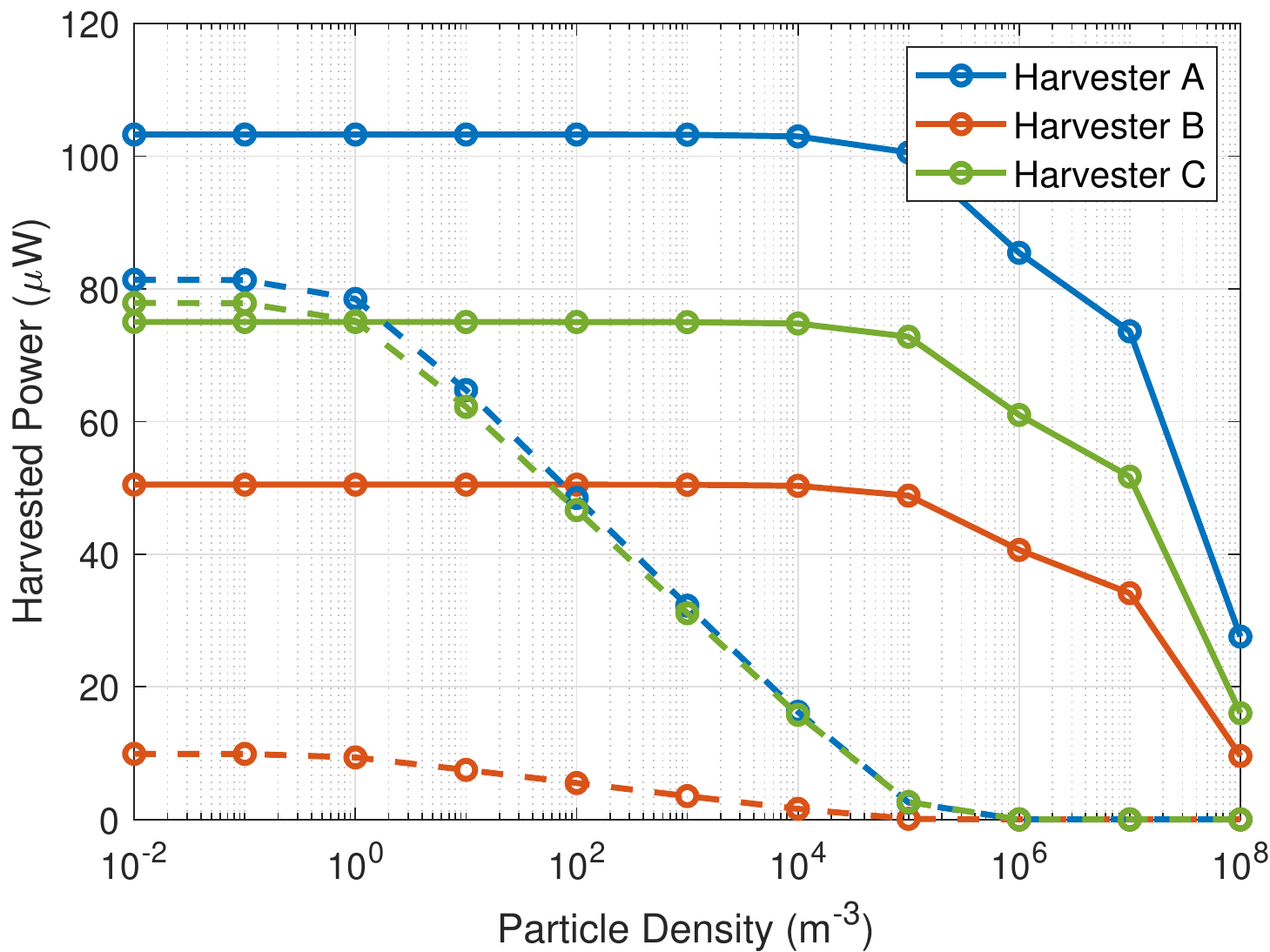}}
\subfigure[]{
\label{fig:particle_density_10W_50m_1e-4_5e-3_dust_1_misalignment_0_area_2}
\includegraphics[width= 0.45\linewidth]{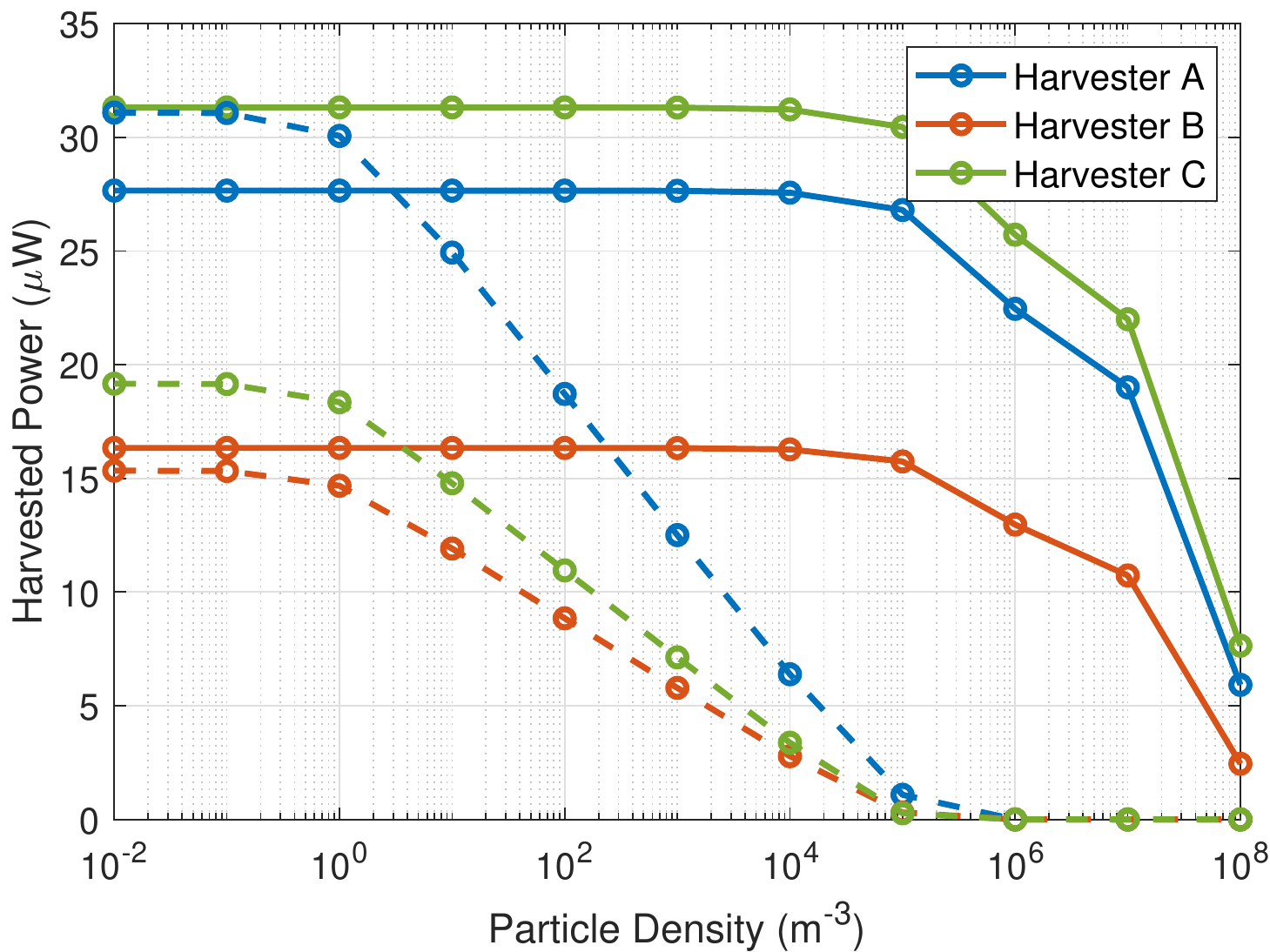}}
\caption{The harvested power by three different harvesters in (a) Gale Crater Area 1 and (b) Gale Crater Area 2 versus dust particle density when the transmit power and distance are 10 W and 50 m, respectively. The solid lines and dashed lines denote the particle sizes of $1\times10^{-4}$ and $5\times10^{-3}$ m, respectively. It should be noted that the particle sizes are selected higher than measured values to show the robustness of RF propagation under dust storms.}
\label{fig:particle_density_10W_50m_1e-4_5e-3_dust_1_misalignment_0}
\end{figure*}

As mentioned above, dust storms in the Martian atmosphere have a degrading effect on the quality of RF propagation, and thus the harvested power decreases according to the dust density and the size of dust particles. The simulation results regarding dust storms are depicted in \FGR{fig:particle_density_10W_50m_1e-4_5e-3_dust_1_misalignment_0}. In the simulation, we investigate the harvested power under the assumption of large particles to show the robustness of RF \ac{WPT} against dust storms. In this simulation, the distance between source and destination is 50 m and the transmit power is 10 W. In both areas of Gale Crater, the Harvester B is underperforming the others. As seen, in case the particle size is 100 $\mu$m, it is seen that there is no significant change in the amount of energy harvested until the particle density increases to $10^5$ m$^{-3}$. As known, particle radius in dust storms on Mars~\cite{lemmon2019martian} is far below the value used in this simulation. This shows that RF \ac{WPT} is more robust than transmission with visible light or laser~\cite{perera2017simultaneous}.

Finally, the effect of the alignment between the receiver and transmitter antennas is investigated. As it is known, it is desired to increase efficiency by using sharp beams in power transmission. In that case, high-resolution channel estimation is required to adjust antenna alignment. But channel estimation is computationally complex. Moreover, since \ac{ZE} devices have limited computational capabilities and simple transceivers, both estimation and alignment pose a challenging task for \ac{ZE} devices.

\begin{figure*}[!t]
\centering
\subfigure[]{
\label{fig:error_jitter_10W_50m_dust_0_misalignment_1_beta_05_1_area_1}
\includegraphics[width= 0.45\linewidth]{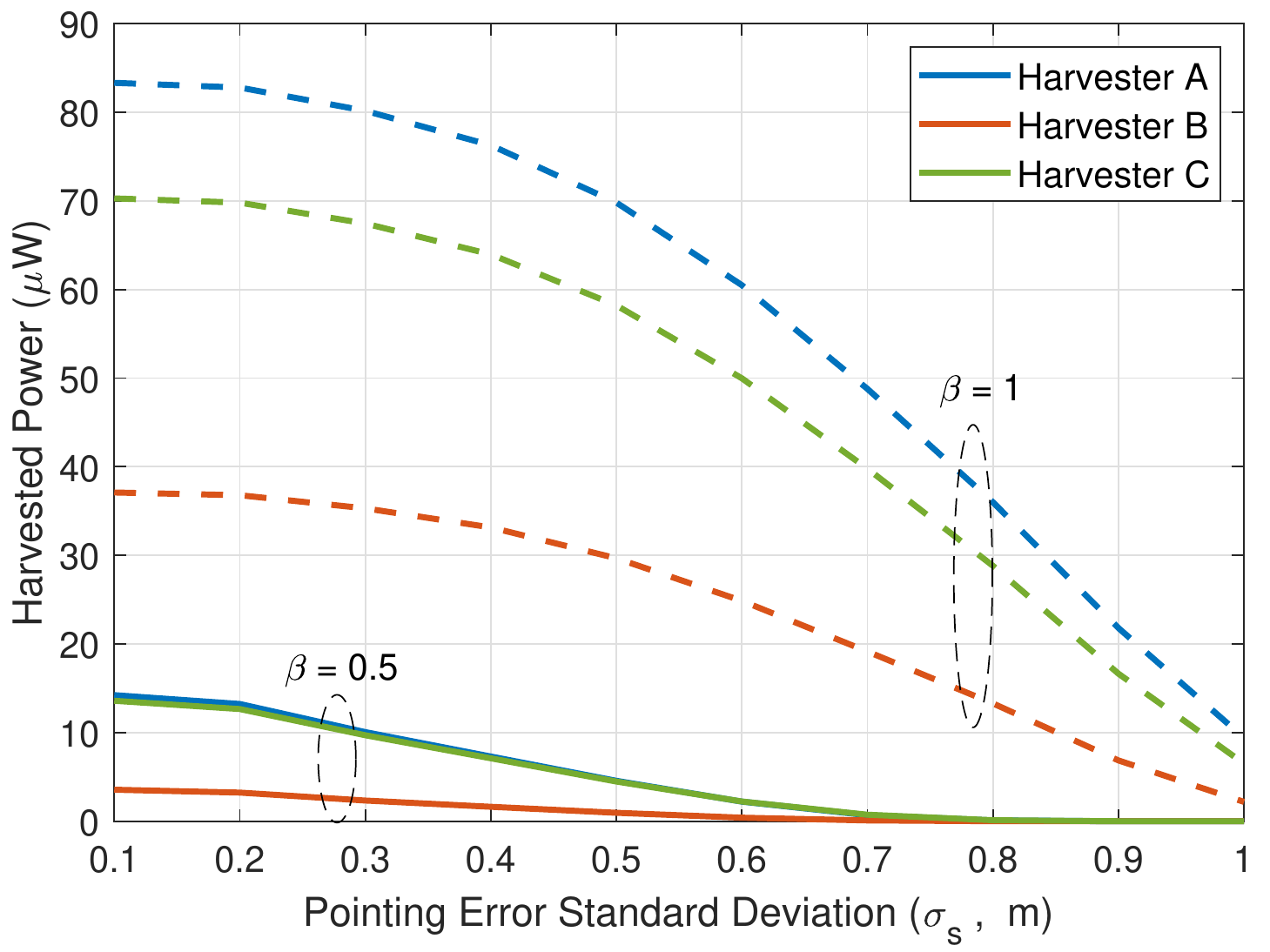}}
\subfigure[]{
\label{fig:error_jitter_10W_50m_dust_0_misalignment_1_beta_05_1_area_2}
\includegraphics[width= 0.45\linewidth]{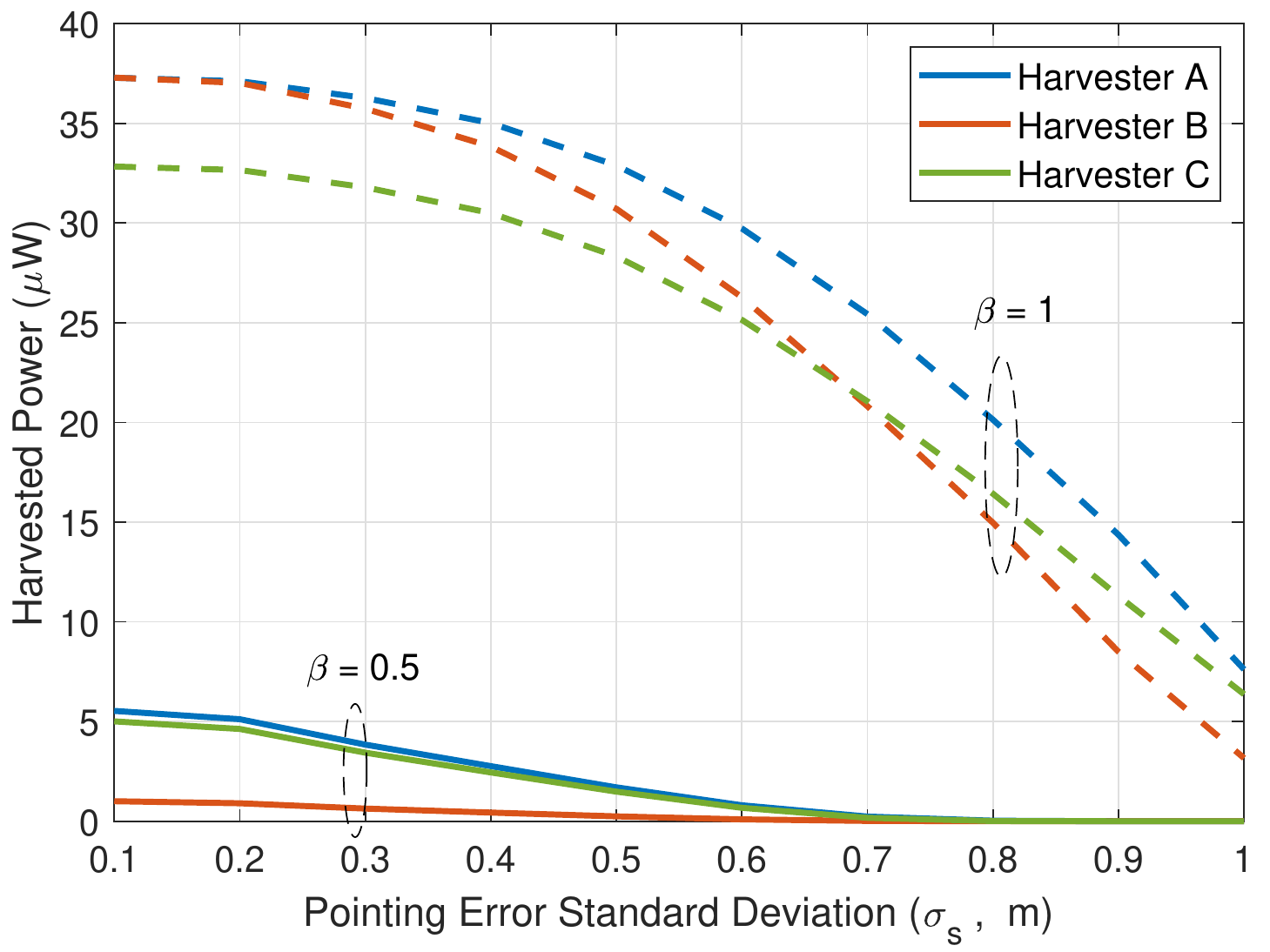}}
\caption{The harvested power by three different harvesters in (a) Gale Crater Area 1 and (b) Gale Crater Area 2 versus the standard deviation of pointing error when the transmit power and distance are 10 W and 50 m, respectively. The solid lines and dashed lines denote the radius of receiver beam aperture of $0.5$ and $1$ m, respectively. It is shown that Area 2 needs both robust alignment and a wider receiver beam aperture.}
\label{fig:error_jitter_10W_50m_dust_0_misalignment_1_beta_05_1}
\end{figure*}

Therefore, some misalignment between the transceiver antennas is expected. Depending on the standard deviation of the pointing error and the radius of the receiver beam aperture, harvested power is simulated for both areas. As expected, increasing misalignment jitter degrades the harvested power. As given in \FGR{fig:error_jitter_10W_50m_dust_0_misalignment_1_beta_05_1}, there is an almost linear relationship between harvested power and pointing error when the standard deviation of pointing error is between 0.4 and 1 m. On the other hand, the power loss owing to misalignment fading can be tolerated by increasing antenna aperture of the receiver. It should be noted that due to the limited antenna capabilities of \ac{ZE} devices, it is obvious that there is a strict limit for aperture size. Harvester B generally shows poor performance compared to its peers. Especially in the case of high pointing error which causes drops in the incident power, the performance decrease is more pronounced. This result is in line with the results depicted in \FGR{fig:harvester_eff}. While it is possible to use devices with relatively smaller antenna apertures in Area 1 of Gale Crater, much larger antenna apertures and/or more robust antenna alignment are needed in Area 2. At this point, a system design problem arises, which requires evaluating resources and constraints to determine a strategy for optimum harvesting.

The long and the short of it is shown that the harvested power by the destination node is above a couple of $\mu$W although under pointing error and dust storms. Considering the numerical results and power requirements of the novel \ac{ZE} devices~\cite{lee2012modular, zhang2018near}, it can be readily said that the power can be supplied for \ac{ZE} devices on the Martian surface by WPT with the conventional and simple harvesters.

\section{Open Issues and Future Directions}\label{sec:open_issues}

Although the results given in the previous section show that \ac{WPT} on the Martian surface is promising for \ac{ZE} devices, there are many open issues that need to be addressed. We discuss some of them in this section. 

\paragraph{Channel Models for Martian Propagation Environment}
In this study, we discussed how the harvested power changes according to distance, transmission power, dust storms, and pointing error. The wireless channel model that we employed in this study is valid for Gale Crater, and there is a need to create channel models for the remaining regions of Mars. Consistent channel models should be obtained so that the results given in this study can be generalized to the whole of Mars. Furthermore, even though some recent papers~\cite{bonafini2020building, pucci2010performance} assume the multipath channel is Rayleigh when considering the small-scale fading in Mars, the research on small-scale characteristics of the Martian propagation medium is scarce. To develop accurate models and investigate the harvesting performance, the mature channel models including small-scale characteristics are strictly needed. Therefore, studies on channel modeling for Mars should be addressed in future scientific publications first. 

\paragraph{Harvester Design and Efficiency Models}
In addition, there may be a need to design novel harvesters for the purpose addressed in this study, not being limited to the three harvester models used above. Also, for novel harvesters to be designed, it may not be possible to express the relation of harvester efficiency with incident power with the heuristic model, or the fitting error for the heuristic model may be high. Therefore, there may be a need for new efficiency models that include the effect of environmental conditions such as temperature.

\paragraph{Non-line-of-sight Wireless Power Transmission}
Although it is assumed that \ac{ZE} devices are located in the transmitter's line-of-sight throughout the study, analyzes and new methods are also required for non-line-of-sight (NLOS) propagation. Harvested power is expected to decrease significantly in NLOS conditions. The initial results of the studies on the reconfigurable intelligent surface-aided \ac{WPT} show that the harvested power amount can be increased compared to the harvested power in NLOS case by properly positioning a reflective surfaces between the source and the destination~\cite{tran2021reconfigurable, yang2021reconfigurable}. Also, the reflective surfaces can focus RF beam through the destination~\cite{ellingson2021path}. 

\paragraph{Battery Recharging Time Analysis}
Because of the sporadic and random nature of the harvested power, it may not be directly utilized by the \ac{ZE} devices~\cite{altinel2014statistical}. In general, the harvested power is stored in rechargeable batteries. Thus, modeling battery recharging time becomes crucial in the system design perspective. Hence, recharging time for batteries suitable for the Mars environment should be statistically modeled in order to determine the transmit power, battery selection, and so on.

\paragraph{Multi-source Harvesting}
Along with RF power transmission, it is possible to design hybrid systems by making use of other energy sources in the Martian environment such as solar, wind, vibration, etc~\cite{altinel2019modeling}. In addition, wireless power can be transmitted from multiple RF sources~\cite{altinel2016energy}. On the other hand, EH from other energy sources in the Martian environment is still an open issue. Although EH from other energy sources can supply more power~\cite{alves2021wireless}, the hardware complexity of harvesters is quite high for \ac{ZE} devices. Moreover, it should be noted that most of the non-RF harvesters have been designed to operate on Earth contrary to what we proposed in this work. It is yet an open issue how such harvesters can operate in the Martian environment and what their efficiency will be. Also, the conformal integrated solar panel antennas~\cite{yekan2017conformal} can pave the way for joint utilization of solar and RF harvesting.

\section{Conclusion}\label{sec:conclusion}

In this study, a preliminary investigation is carried out to provide the power required by \ac{ZE} devices with \ac{WPT} under the environmental conditions of Mars. First, the efficiencies of different harvesters designed recently are modeled and RF path loss models are discussed in the Mars propagation environment. Furthermore, the effect of dust storms on Mars and beam misalignment between the source and destinations on harvested power is also being studied. The initial results show that the power required for the proper operation of \ac{ZE} devices can be provided without employing any sophisticated hardware. In addition, the open issues in this study are detailed and the direction for future studies is provided.

\bibliographystyle{IEEEtran}
\bibliography{IEEEabrv, main}

\begin{thebibliography}{10}
\providecommand{\url}[1]{#1}
\csname url@samestyle\endcsname
\providecommand{\newblock}{\relax}
\providecommand{\bibinfo}[2]{#2}
\providecommand{\BIBentrySTDinterwordspacing}{\spaceskip=0pt\relax}
\providecommand{\BIBentryALTinterwordstretchfactor}{4}
\providecommand{\BIBentryALTinterwordspacing}{\spaceskip=\fontdimen2\font plus
\BIBentryALTinterwordstretchfactor\fontdimen3\font minus
  \fontdimen4\font\relax}
\providecommand{\BIBforeignlanguage}[2]{{%
\expandafter\ifx\csname l@#1\endcsname\relax
\typeout{** WARNING: IEEEtran.bst: No hyphenation pattern has been}%
\typeout{** loaded for the language `#1'. Using the pattern for}%
\typeout{** the default language instead.}%
\else
\language=\csname l@#1\endcsname
\fi
#2}}
\providecommand{\BIBdecl}{\relax}
\BIBdecl

\bibitem{MissionO34:online}
NASA, ``{Mission Overview - NASA Mars},''
  \url{https://mars.nasa.gov/mars2020/mission/overview/}, (Accessed on
  11/17/2021).

\bibitem{barnes2020multiple}
J.~J. Barnes, F.~M. McCubbin, A.~R. Santos, J.~M. Day, J.~W. Boyce, S.~P.
  Schwenzer, U.~Ott, I.~A. Franchi, S.~Messenger, M.~Anand \emph{et~al.},
  ``{Multiple early-formed water reservoirs in the interior of Mars},''
  \emph{Nature Geoscience}, vol.~13, no.~4, pp. 260--264, 2020.

\bibitem{bonafini2020building}
S.~Bonafini and C.~Sacchi, ``{Building cellular connectivity on Mars: A
  feasibility study},'' in \emph{IEEE Aerospace Conference}, 2020, pp. 1--12.

\bibitem{alazzam2011thermal}
A.~Alazzam, L.~Ngo~Phong, and M.~Daly, ``A thermal anemometer for the {Mars}
  meteorological sensor network,'' in \emph{AIAA SPACE Conference \&
  Exposition}, 2011, p. 7317.

\bibitem{lopez2021massive}
O.~L. L{\'o}pez, H.~Alves, R.~D. Souza, S.~Montejo-S{\'a}nchez, E.~M.~G.
  Fern{\'a}ndez, and M.~Latva-Aho, ``{Massive wireless energy transfer:
  Enabling sustainable IoT toward 6G era},'' \emph{{IEEE} Internet Things J.},
  vol.~8, no.~11, pp. 8816--8835, 2021.

\bibitem{lee2012modular}
Y.~Lee, S.~Bang, I.~Lee, Y.~Kim, G.~Kim, M.~H. Ghaed, P.~Pannuto, P.~Dutta,
  D.~Sylvester, and D.~Blaauw, ``{A modular 1 mm$^3$ die-stacked sensing
  platform with low power I$^2$C inter-die communication and multi-modal energy
  harvesting},'' \emph{{IEEE} J. Solid-State Circuits}, vol.~48, no.~1, pp.
  229--243, 2012.

\bibitem{haque2020supplemental}
T.~Haque, H.~Elkotby, P.~Cabrol, R.~Pragada, and D.~Castor, ``{A supplemental
  zero-energy downlink air-interface enabling 40-year battery life in IoT
  devices},'' in \emph{IEEE Global Communications Conference}, 2020, pp. 1--6.

\bibitem{alves2021wireless}
H.~Alves and O.~A. L{\'o}pez, \emph{{Wireless RF Energy Transfer in the Massive
  IoT Era}}.\hskip 1em plus 0.5em minus 0.4em\relax John Wiley \& Sons, 2021.

\bibitem{taha2021eliminating}
A.~Taha, H.~Elkotby, T.~Haque, R.~Pragada, and D.~Castor, ``Eliminating battery
  replacement throughout the useful life of {IoT} devices with limited-capacity
  batteries: {Analysis} and design of a zero energy air interface,'' in
  \emph{IEEE International Conference on Communications Workshops}, 2021, pp.
  1--6.

\bibitem{zhang2018near}
R.~Zhang, S.~Fan, and L.~Geng, ``{A near-zero-power temperature sensor
  with$\pm$0.24° C inaccuracy using only standard CMOS transistors for IoT
  applications},'' in \emph{IEEE International Symposium on Circuits and
  Systems (ISCAS)}, 2018, pp. 1--4.

\bibitem{portilla2019extreme}
J.~Portilla, G.~Mujica, J.-S. Lee, and T.~Riesgo, ``{The extreme edge at the
  bottom of the Internet of Things: A review},'' \emph{{IEEE} Sensors J.},
  vol.~19, no.~9, pp. 3179--3190, 2019.

\bibitem{mahmood2020white}
N.~H. Mahmood, S.~B{\"o}cker, A.~Munari, F.~Clazzer, I.~Moerman, K.~Mikhaylov,
  O.~Lopez, O.-S. Park, E.~Mercier, H.~Bartz \emph{et~al.}, ``{White paper on
  critical and massive machine type communication towards 6G},'' \emph{arXiv
  preprint arXiv:2004.14146}, 2020.

\bibitem{mahmood2020six}
N.~H. Mahmood, H.~Alves, O.~A. L{\'o}pez, M.~Shehab, D.~P.~M. Osorio, and
  M.~Latva-Aho, ``{Six key features of machine type communication in 6G},'' in
  \emph{2nd 6G Wireless Summit (6G SUMMIT)}, 2020, pp. 1--5.

\bibitem{sherman2021design}
M.~Sherman and M.~Hassanalian, ``{Design, fabrication, and testing of
  dandelion-inspired flying sensors for Mars exploration},'' in \emph{AIAA
  Scitech 2021 Forum}, 2021, p. 0962.

\bibitem{dardari2019ultra}
D.~Dardari, N.~Decarli, D.~Fabbri, A.~Guerra, M.~Fantuzzi, D.~Masotti,
  A.~Costanzo, A.~Romani, M.~Drouguet, T.~Feuillen \emph{et~al.}, ``An
  ultra-wideband battery-less positioning system for space applications,'' in
  \emph{IEEE International Conference on RFID Technology and Applications},
  2019, pp. 104--109.

\bibitem{sasaki2014japan}
S.~Sasaki, ``{How Japan plans to build an orbital solar farm},'' \emph{IEEE
  Spectrum}, vol.~24, pp. 46--51, 2014.

\bibitem{fuse2011outline}
Y.~Fuse, T.~Saito, S.~Mihara, K.~Ijichi, K.~Namura, Y.~Honma, T.~Sasaki,
  Y.~Ozawa, E.~Fujiwara, and T.~Fujiwara, ``{Outline and progress of the
  Japanese microwave energy transmission program for SSPS},'' in \emph{2011
  IEEE MTT-S International Microwave Workshop Series on Innovative Wireless
  Power Transmission: Technologies, Systems, and Applications}, 2011, pp.
  47--50.

\bibitem{fuse2011microwave}
------, ``{Microwave energy transmission program for SSPS},'' in \emph{30th
  URSI General Assembly and Scientific Symposium}, 2011, pp. 1--4.

\bibitem{936Scott70:online}
J.~H. Scott, ``{STMD}’s power and energy storage roadmap and gap closure
  plan: Wireless power transfer,'' \url{https://bit.ly/3noBDhf}, July 2021,
  (Accessed on 11/18/2021).

\bibitem{liu2015multiple}
C.~Liu, K.~Chau, Z.~Zhang, C.~Qiu, F.~Lin, and T.~Ching, ``Multiple-receptor
  wireless power transfer for magnetic sensors charging on mars via magnetic
  resonant coupling,'' \emph{Journal of Applied Physics}, vol. 117, no.~17, p.
  17A743, 2015.

\bibitem{landis2000solar}
G.~A. Landis, ``{Solar cell selection for Mars},'' \emph{IEEE Aerosp. Electron.
  Syst. Mag.}, vol.~15, no.~1, pp. 17--21, 2000.

\bibitem{kaplan1988environment}
D.~Kaplan, \emph{{Environment of Mars}}.\hskip 1em plus 0.5em minus 0.4em\relax
  National Aeronautics and Space Administration, 1988, vol. 100470.

\bibitem{appelbaum1990solar}
J.~Appelbaum and D.~J. Flood, ``{Solar radiation on Mars},'' \emph{Solar
  Energy}, vol.~45, no.~6, pp. 353--363, 1990.

\bibitem{haberle1993atmospheric}
R.~M. Haberle, C.~P. McKay, J.~Pollack, O.~Gwynne, D.~Atkinson, J.~Appelbaum,
  G.~Landis, R.~Zurek, and D.~Flood, ``{Atmospheric effects on the utility of
  solar power on Mars},'' \emph{Resources of Near-Earth Space}, p. 845, 1993.

\bibitem{hess1977meteorological}
S.~Hess, R.~Henry, C.~B. Leovy, J.~Ryan, and J.~E. Tillman, ``{Meteorological
  results from the surface of Mars: Viking 1 and 2},'' \emph{Journal of
  Geophysical Research}, vol.~82, no.~28, pp. 4559--4574, 1977.

\bibitem{landis2004mars}
G.~Landis, T.~Kerslake, D.~Scheiman, and P.~Jenkins, ``Mars solar power,'' in
  \emph{2nd International Energy Conversion Engineering Conference}, 2004, p.
  5555.

\bibitem{mckissock1990solar}
B.~I. McKissock, L.~L. Kohout, and P.~C. Schmitz, ``{A solar power system for
  an early Mars expedition},'' in \emph{American Institute of Chemical
  Engineers Summer National Meeting}, no. E-5632, 1990.

\bibitem{kerslake1999solar}
T.~W. Kerslake and L.~L. Kohout, ``{Solar electric power system analyses for
  Mars surface missions},'' SAE Technical Paper, Tech. Rep., 1999.

\bibitem{ortabasi2006powersphere}
U.~Ortabasi and H.~Friedman, ``{Powersphere: A photovoltaic cavity converter
  for wireless power transmission using high power lasers},'' in \emph{IEEE 4th
  World Conference on Photovoltaic Energy Conference}, vol.~1, 2006, pp.
  126--129.

\bibitem{iwashimizu2014study}
M.~Iwashimizu, T.~Mitani, N.~Shinohara, G.~Sasaki, K.~Hiraoka, K.~Matsuzaki,
  and K.~Yonemoto, ``{Study on direction detection in a microwave power
  transmission system for a Mars observation airplane},'' in \emph{IEEE
  Wireless Power Transfer Conference}, 2014, pp. 146--149.

\bibitem{xie2013wireless}
L.~Xie, Y.~Shi, Y.~T. Hou, and A.~Lou, ``Wireless power transfer and
  applications to sensor networks,'' \emph{IEEE Wirel. Commun.}, vol.~20,
  no.~4, pp. 140--145, 2013.

\bibitem{chukkala2005simulation}
V.~Chukkala and P.~De~Leon, ``{Simulation and analysis of the multipath
  environment of Mars},'' in \emph{IEEE Aerospace Conference}, 2005, pp.
  1678--1683.

\bibitem{daga2007terrain}
A.~Daga, G.~R. Lovelace, D.~K. Borah, and P.~L. De~Leon, ``{Terrain-based
  simulation of IEEE 802.11 a and b physical layers on the martian surface},''
  \emph{IEEE Trans. Aerosp. Electron. Syst.}, vol.~43, no.~4, pp. 1617--1624,
  2007.

\bibitem{del2009ieee802}
E.~Del~Re, R.~Pucci, and L.~S. Ronga, ``{IEEE802. 15.4 wireless sensor network
  in Mars exploration scenario},'' in \emph{International Workshop on Satellite
  and Space Communications}, 2009, pp. 284--288.

\bibitem{bonafini2020evaluation}
S.~Bonafini and C.~Sacchi, ``Evaluation of large scale propagation phenomena on
  the {Martian} surface: a {3D} ray tracing approach,'' in \emph{10th Advanced
  Satellite Multimedia Systems Conference and the 16th Signal Processing for
  Space Communications Workshop}, 2020, pp. 1--8.

\bibitem{wray2013gale}
J.~J. Wray, ``{Gale crater: the Mars Science Laboratory/Curiosity rover landing
  site},'' \emph{International Journal of Astrobiology}, vol.~12, no.~1, pp.
  25--38, 2013.

\bibitem{hassler2014mars}
D.~M. Hassler, C.~Zeitlin, R.~F. Wimmer-Schweingruber, B.~Ehresmann, S.~Rafkin,
  J.~L. Eigenbrode, D.~E. Brinza, G.~Weigle, S.~B{\"o}ttcher, E.~B{\"o}hm
  \emph{et~al.}, ``{Mars’ surface radiation environment measured with the
  Mars Science Laboratory’s Curiosity rover},'' \emph{science}, vol. 343, no.
  6169, p. 1244797, 2014.

\bibitem{mahaffy2013abundance}
P.~R. Mahaffy, C.~R. Webster, S.~K. Atreya, H.~Franz, M.~Wong, P.~G. Conrad,
  D.~Harpold, J.~J. Jones, L.~A. Leshin, H.~Manning \emph{et~al.}, ``{Abundance
  and isotopic composition of gases in the Martian atmosphere from the
  Curiosity rover},'' \emph{Science}, vol. 341, no. 6143, pp. 263--266, 2013.

\bibitem{goldhirsh1982parameter}
J.~Goldhirsh, ``{A parameter review and assessment of attenuation and
  backscatter properties associated with dust storms over desert regions in the
  frequency range of 1 to 10 GHz},'' \emph{IEEE Trans. Antennas Propag.},
  vol.~30, no.~6, pp. 1121--1127, 1982.

\bibitem{sacchi2019lte}
C.~Sacchi and S.~Bonafini, ``{From LTE-A to LTE-M: A futuristic convergence
  between terrestrial and Martian mobile communications},'' in \emph{IEEE
  International Black Sea Conference on Communications and Networking
  (BlackSeaCom)}, 2019, pp. 1--5.

\bibitem{farid_outage_2007}
A.~A. Farid and S.~Hranilovic, ``Outage capacity optimization for free-space
  optical links with pointing errors,'' \emph{Journal of Lightwave Technology},
  vol.~25, no.~7, pp. 1702--1710, Jul. 2007.

\bibitem{boulogeorgos_error_2020}
A.-A.~A. Boulogeorgos and A.~Alexiou, ``Error analysis of mixed {THz}-{RF}
  wireless systems,'' \emph{IEEE Commun. Lett.}, vol.~24, no.~2, pp. 277--281,
  Feb. 2020.

\bibitem{cansiz2019efficiency}
M.~Cansiz, D.~Altinel, and G.~K. Kurt, ``{Efficiency in RF energy harvesting
  systems: A comprehensive review},'' \emph{Energy}, vol. 174, pp. 292--309,
  2019.

\bibitem{ju2013throughput}
H.~Ju and R.~Zhang, ``Throughput maximization in wireless powered communication
  networks,'' \emph{IEEE Trans. Wirel. Commun.}, vol.~13, no.~1, pp. 418--428,
  2013.

\bibitem{chen2016new}
Y.~Chen, K.~T. Sabnis, and R.~A. Abd-Alhameed, ``{New formula for conversion
  efficiency of RF EH and its wireless applications},'' \emph{IEEE Trans. Veh.
  Technol.}, vol.~65, no.~11, pp. 9410--9414, 2016.

\bibitem{valenta2014harvesting}
C.~R. Valenta and G.~D. Durgin, ``{Harvesting wireless power: Survey of
  energy-harvester conversion efficiency in far-field, wireless power transfer
  systems},'' \emph{IEEE Microw. Mag.}, vol.~15, no.~4, pp. 108--120, 2014.

\bibitem{clerckx2018fundamentals}
B.~Clerckx, R.~Zhang, R.~Schober, D.~W.~K. Ng, D.~I. Kim, and H.~V. Poor,
  ``{Fundamentals of wireless information and power transfer: From RF energy
  harvester models to signal and system designs},'' \emph{IEEE J. Sel. Areas
  Commun.}, vol.~37, no.~1, pp. 4--33, 2018.

\bibitem{franciscatto2013high}
B.~R. Franciscatto, V.~Freitas, J.-M. Duchamp, C.~Defay, and T.~P. Vuong,
  ``{High-efficiency rectifier circuit at 2.45 GHz for low-input-power RF
  energy harvesting},'' in \emph{European Microwave Conference}, 2013, pp.
  507--510.

\bibitem{zhang2016high}
X.~Y. Zhang, Z.-X. Du, and Q.~Xue, ``High-efficiency broadband rectifier with
  wide ranges of input power and output load based on branch-line coupler,''
  \emph{IEEE Transactions on Circuits and Systems}, vol.~64, no.~3, pp.
  731--739, 2016.

\bibitem{lau2020deep}
W.~W.~Y. Lau, H.~W. Ho, and L.~Siek, ``{Deep neural network (DNN) optimized
  design of 2.45 GHz CMOS rectifier with 73.6\% peak efficiency for RF energy
  harvesting},'' \emph{IEEE Transactions on Circuits and Systems}, vol.~67,
  no.~12, pp. 4322--4333, 2020.

\bibitem{ramalingam2021advancement}
L.~Ramalingam, S.~Mariappan, P.~Parameswaran, J.~Rajendran, R.~S. Nitesh,
  N.~Kumar, A.~Nathan, and B.~S. Yarman, ``The advancement of radio frequency
  energy harvesters {(RFEHs)} as a revolutionary approach for solving energy
  crisis in wireless communication devices: {A} review,'' \emph{IEEE Access},
  2021.

\bibitem{liu2021research}
W.~Liu, Y.~Wang, and J.~Song, ``{Research on Schottky diode with high
  rectification efficiency for relatively weak energy wireless harvesting},''
  \emph{Superlattices and Microstructures}, vol. 150, p. 106639, 2021.

\bibitem{lemmon2019martian}
M.~Lemmon, S.~Guzewich, T.~McConnochie, G.~Mart{\'\i}nez,
  {\'A}.~de~Vicente-Retortillo, M.~Smith, J.~Bell, D.~Wellington, and
  S.~Jacobs, ``Martian dust particle size during the 2018 planet-encircling
  dust storm as measured by the {Curiosity Rover},'' \emph{LPI Contributions},
  vol. 2089, p. 6298, 2019.

\bibitem{perera2017simultaneous}
T.~D.~P. Perera, D.~N.~K. Jayakody, S.~K. Sharma, S.~Chatzinotas, and J.~Li,
  ``{Simultaneous wireless information and power transfer (SWIPT): Recent
  advances and future challenges},'' \emph{IEEE Commun. Surv. Tutor.}, vol.~20,
  no.~1, pp. 264--302, 2017.

\bibitem{pucci2010performance}
R.~Pucci, E.~D. Re, D.~Boschetti, and L.~Ronga, ``{Performance evaluation of an
  IEEE802. 15.4 standard based wireless sensor network in Mars exploration
  scenario},'' in \emph{The Internet of Things}.\hskip 1em plus 0.5em minus
  0.4em\relax Springer, 2010, pp. 349--358.

\bibitem{tran2021reconfigurable}
N.~M. Tran, M.~M. Amri, J.~H. Park, D.~I. Kim, and K.~W. Choi, ``Reconfigurable
  intelligent surface-aided wireless power transfer systems: {Analysis} and
  implementation,'' \emph{arXiv preprint arXiv:2106.11805}, 2021.

\bibitem{yang2021reconfigurable}
H.~Yang, X.~Yuan, J.~Fang, and Y.-C. Liang, ``Reconfigurable intelligent
  surface aided constant-envelope wireless power transfer,'' \emph{IEEE Trans.
  Signal Process.}, vol.~69, pp. 1347--1361, 2021.

\bibitem{ellingson2021path}
S.~W. Ellingson, ``Path loss in reconfigurable intelligent surface-enabled
  channels,'' in \emph{IEEE 32nd Annual International Symposium on Personal,
  Indoor and Mobile Radio Communications}, 2021, pp. 829--835.

\bibitem{altinel2014statistical}
D.~Altinel and G.~K. Kurt, ``Statistical models for battery recharging time in
  {RF} energy harvesting systems,'' in \emph{IEEE Wireless Communications and
  Networking Conference (WCNC)}, 2014, pp. 636--641.

\bibitem{altinel2019modeling}
------, ``{Modeling of multiple energy sources for hybrid energy harvesting IoT
  systems},'' \emph{IEEE Internet Things J.}, vol.~6, no.~6, pp.
  10\,846--10\,854, 2019.

\bibitem{altinel2016energy}
------, ``{Energy harvesting from multiple RF sources in wireless fading
  channels},'' \emph{IEEE Trans. Veh. Technol.}, vol.~65, no.~11, pp.
  8854--8864, 2016.

\bibitem{yekan2017conformal}
T.~Yekan and R.~Baktur, ``Conformal integrated solar panel antennas: {Two}
  effective integration methods of antennas with solar cells,'' \emph{IEEE
  Antennas Propag. Mag.}, vol.~59, no.~2, pp. 69--78, 2017.

\end{thebibliography}

\begin{IEEEbiography}
[{\includegraphics[width=1in,height=1.25in,clip,keepaspectratio]{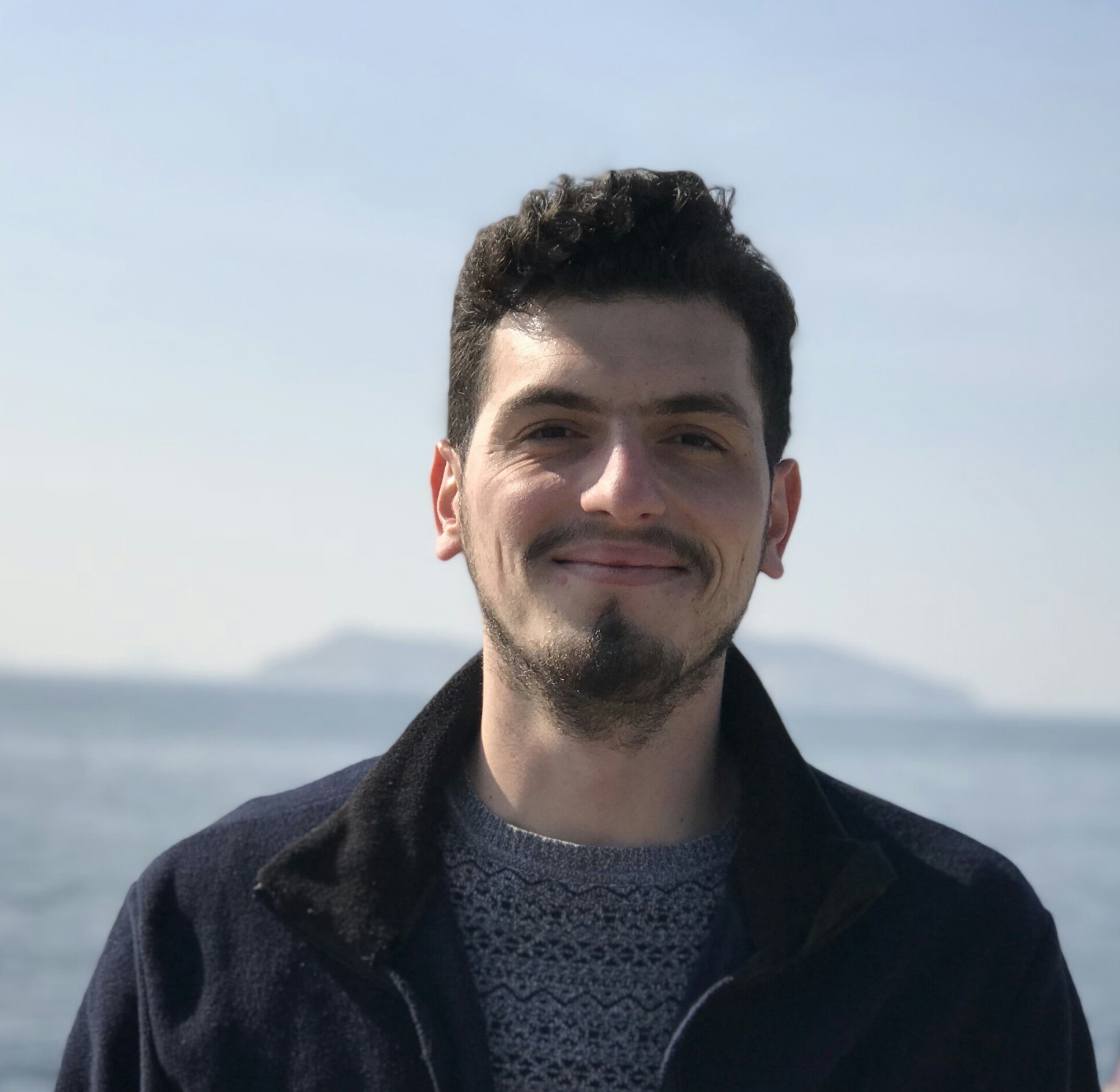}}]{K{\"{u}}r{\c{s}}at Tekb{\i}y{\i}k} received his B.Sc. and M.Sc. degrees (with high honors) in electronics and communication engineering from Istanbul Technical University, Istanbul, Turkey, in 2017 and 2019, respectively. He is currently pursuing his Ph.D. degree in telecommunications engineering at Istanbul Technical University. He is also working as wireless systems and machine learning engineer in a fast-growing Silicon Valley company. His research interests include deep learning applications in wireless communications, terahertz communications, non-terrestrial networks, and reconfigurable intelligent surfaces.
\end{IEEEbiography}

\begin{IEEEbiography}
[{\includegraphics[width=1in,height=1.25in,clip,keepaspectratio]{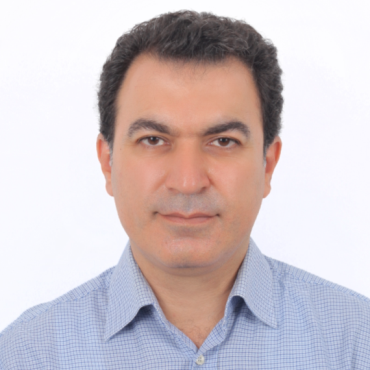}}]{Dogay Altinel} received the B.S. degree in electronics engineering from Hacettepe University, Ankara, Turkey, in 1992, and the M.Sc. and Ph.D. degrees in telecommunications engineering from Istanbul Technical University, Istanbul, Turkey, in 2014 and 2019, respectively. He worked in the electronics and telecommunication industry for many years from 1992 to 2012. He has been working at Istanbul Medeniyet University, Istanbul, since 2012. His research interests include the areas of communication theory, wireless networks, and energy harvesting systems.
\end{IEEEbiography}

\begin{IEEEbiography}
[{\includegraphics[width=1in,height=1.25in,clip,keepaspectratio]{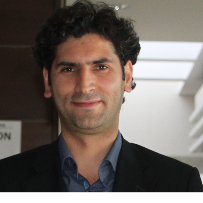}}]{Mustafa Cansiz} received the B.S. degree in electrical and electronics engineering from Karadeniz Technical University, Trabzon, Turkey, in 2002, the M.Sc. degree in electrical and electronics engineering from Dicle University, Diyarbakır, Turkey, in 2010, and the Ph.D. degree in electrical and electronics engineering from Inonu University, Malatya, Turkey, in 2016. From 2005 to 2011, he worked in the electronics and telecommunication industry. Since 2011, he has been with Dicle University. His research areas include the measurement of electromagnetic exposure and radio-frequency energy-harvesting systems.
\end{IEEEbiography}

\begin{IEEEbiography}
[{\includegraphics[width=1in,height=1.25in,clip,keepaspectratio]{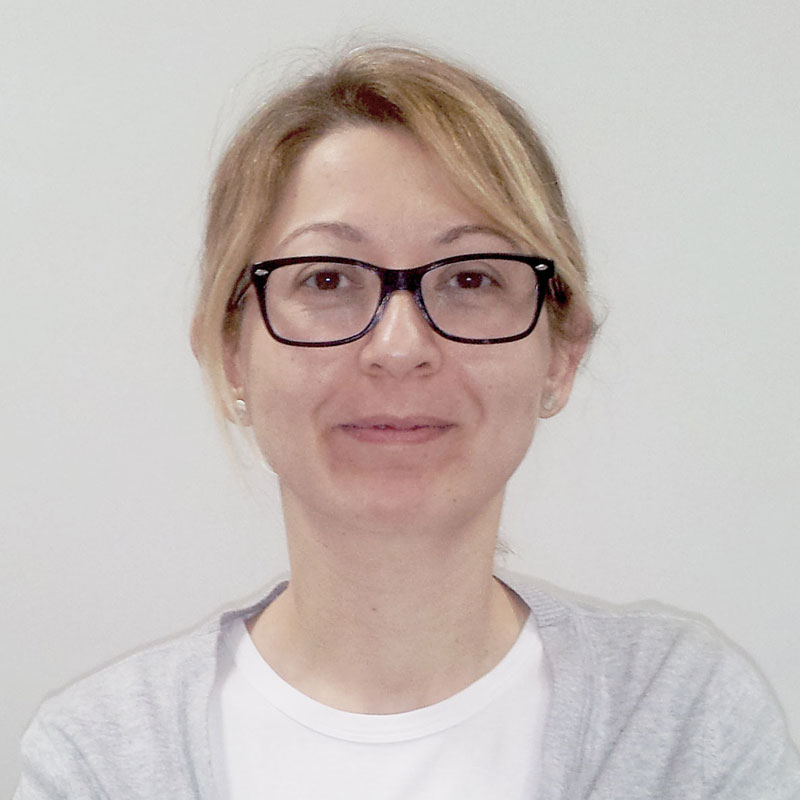}}]{Güneş Karabulut Kurt}
is currently an Associate Professor of Electrical Engineering at Polytechnique Montr\'eal, Montr\'eal, QC, Canada. She received the B.S. degree with high honors in electronics and electrical engineering from the Bogazici University, Istanbul, Turkey, in 2000 and the M.A.Sc. and the Ph.D. degrees in electrical engineering from the University of Ottawa, ON, Canada, in 2002 and 2006, respectively. From 2000 to 2005, she was a Research Assistant at the University of Ottawa. Between 2005 and 2006, Gunes was with TenXc Wireless, Canada. From 2006 to 2008, she was with Edgewater Computer Systems Inc., Canada. From 2008 to 2010, she was with Turkcell Research and Development Applied Research and Technology, Istanbul. Gunes was with Istanbul Technical University between 2010 and 2021. She is a Marie Curie Fellow and has received the Turkish Academy of Sciences Outstanding Young Scientist (TÜBA-GEBIP) Award in 2019. She is an Adjunct Research Professor at Carleton University. She is also currently serving as an Associate Technical Editor (ATE) of the IEEE Communications Magazine and a member of the IEEE WCNC Steering Board.
\end{IEEEbiography}


\end{document}